\documentclass[twocolumn]{aastex631}
\usepackage{graphicx}
\shorttitle{Climbing the Cliffs}
\shortauthors{Crompvoets et al.}

\def \totobjs {19~497}
\def \totav {208} 
\def \ysoav {23}%
\def \contav {185}%
\def \ysowomiss {21}
\def \contwomiss {165}
\def \ysotrain {20} 
\def \conttrain {{147}} 
\def \tottestprf {{41}} 
\def \ysotest {{1}}
\def \conttest {{18}}
\def \totadtest {{22}} 
\def \ysoadtest {{2}} 
\def \contadtest{{20}} 

\def \totysos {{685}} 
\def \totysosfrac {{450}} 
\def \finaltotysos {{450}} 

\def \fracsubs {{72\%}}

\def \totprfysos {{256}} 

\def \totprfrfysosfrac {{129}}
\def \totprfnotrfysosfrac {{65}}

\def \totnewysoscompspitz {{427}} 

\def \totnewysos {{413}}
\def \finaltotnewysos {{413}} 
\def \totconfirmedysos {{34}}

\begin{document}
\title{Climbing the Cliffs: Classifying YSOs in the Cosmic Cliffs JWST Data using a Probabilistic Random Forest}

\correspondingauthor{B. L. Crompvoets}
\email{bcrompvoets@uvic.ca}

\author[0000-0001-8900-5550]{B. L. Crompvoets}
\affiliation{Department of Physics and Astronomy, University of Victoria, Victoria, BC, Canada}
\affiliation{NRC Herzberg Astronomy and Astrophysics, 5071 West Saanich Road, Victoria, BC V9E 2E7, Canada}

\author[0000-0002-9289-2450]{J. Di Francesco}
\affiliation{NRC Herzberg Astronomy and Astrophysics, 5071 West Saanich Road, Victoria, BC V9E 2E7, Canada}
\affiliation{Department of Physics and Astronomy, University of Victoria, Victoria, BC, Canada}

\author[0000-0002-4348-6974]{H.Teimoorinia}
\affiliation{NRC Herzberg Astronomy and Astrophysics, 5071 West Saanich Road, Victoria, BC V9E 2E7, Canada}
\affiliation{Department of Physics and Astronomy, University of Victoria, Victoria, BC, Canada}

\author[0000-0003-3130-7796]{T. Preibisch}
\affiliation{Universitäts-Sternwarte München, Ludwig-Maximilians-Universität, Scheinerstr. 1, 81679 München, Germany}

\begin{abstract}
Among the first observations released to the public from the James Webb Space Telescope (JWST) was a section of the star-forming region NGC 3324 known colloquially as the ``Cosmic Cliffs.'' We build a photometric catalog of the region and test the ability of using the Probabilistic Random Forest machine learning method to identify its Young Stellar Objects (YSOs).
We find \finaltotysos{} candidate YSOs (cYSOs) out of \totobjs{} total objects within the field, \finaltotnewysos{} of which are cYSOs not found in previous works. These classifications are verified with several different metrics, including recall and precision. Using the obtained probabilities of objects being YSOs, we employ a Monte Carlo approach to determine the surface density of cYSOs in the Cosmic Cliffs, which we find to be largely coincident with column densities derived from Herschel data, up to a column density of 1.37 $\times$ 10$^{22}$ cm$^{-2}$. The newly determined number and spatial distribution of YSOs in the Cosmic Cliffs demonstrate that JWST is far more capable of detecting YSOs in dusty regions than Spitzer. Comparisons of the observed colors and brightness of faint cYSOs with those of pre-main-sequence models suggest JWST has detected a significant population of sub-stellar YSOs in the Cosmic Cliffs.  The size of this population further suggests previous estimates of star formation efficiencies in molecular clouds have been systematically low.
\end{abstract}

\section{Introduction}

{Stars form within molecular clouds throughout the Galaxy.  The efficiency of such clouds in making stars can be measured by taking the ratio of the total mass of stars produced over a given time to the total mass of the cloud that produced them.  The former quantity can be obtained through censuses of the Young Stellar Objects (YSOs) associated with clouds.  Given that many clouds enshroud their YSOs behind significant amounts of dust intermixed with their gas, sensitive multi-band imaging  of clouds in the infrared is critical for detecting YSO populations.  Moreover, YSOs may be spatially crowded, especially in distant clouds, requiring high resolution observations to identify them individually.  
In this paper, we explore the YSO population revealed by infrared images of the ``Cosmic Cliffs," part of the NGC 3324 star-forming region observed early on by the James Webb Space Telescope (JWST)  utilizing the NIRCam instrument, and whose data were released 
publicly in July 2022.}

Prior to Webb, the infrared Spitzer Space Telescope contributed much to the field of star formation, providing in-depth photometric surveys of several Galactic star-forming regions \citep[e.g.,][]{Majewski2007,Evansc2d}. With these surveys, YSOs in different stages of pre-MS evolution were identified through a variety of techniques \citep[e.g.,][]{Gutermuth2009, Dunham2015}. Even after Spitzer ceased operations, its data continue to be reviewed and more YSOs identified \citep[e.g.,][]{Kuhn2021}.

{Despite the continued usefulness of its data, Spitzer nonetheless had intrinsic sensitivity and resolution limitations in its ability to detect embedded YSOs. 
Fortunately, JWST now surpasses the sensitivity of Spitzer by approximately two orders of magnitude \citep{Rigby2023} in the same field, at a resolution over seven times that of Spitzer at wavelengths in common. 
With JWST operating in both the near- and mid-infrared, it is poised to provide sensitive observations of previously undetected YSOs buried within molecular clouds.}

{Astronomers have recently begun turning to machine learning (ML) as a tool for analyzing large amounts of astronomical data, and indeed ML has been used previously for the classification of YSOs using Spitzer data \citep[e.g., see][]{Cornu2021,Kuhn2021}.
With JWST providing multi-band images with far more sources in small fields than previous telescopes, ML is a logical choice for the rapid 
classification of those sources.
In particular, we examine in this paper the utility of identifying YSOs in the JWST fields using the flexible Probabilistic Random Forest (PRF) approach with the public Cosmic Cliffs data.}
{This paper will also demonstrate the improvements of using JWST NIRCam data over Spitzer data for identifying YSOs. 
Though the Cosmic Cliffs data are relatively limited, further JWST data will allow for the PRF approach to be expanded and refined in the near future.}

{This paper is split into the following sections. 
Section~\ref{sec:Background} summarizes background information relevant to topics described in the paper.
Section~\ref{sec:Method} describes how we created the catalog of JWST data for the Cosmic Cliffs, as well as a description of the PRF method. Section~\ref{sec:Results} provides the results of the PRF-determined classifications, including the number of candidate YSOs identified and a comparison with those found in previous works. 
Section~\ref{sec:Disc} discusses the accuracy of our classifications, and provides an analysis of the implications of the newly identified YSO population.  A summary in Section~\ref{sec:conclude} concludes the paper.}

\section{Background}\label{sec:Background}
{To begin, we provide some background information for this paper.  Below, we describe previous observations of the NGC 3324/Gum 31 region (\S\ref{sec:target_area}) and provide a brief summary of ML tools (\S\ref{sec:ml_tool_bg}). Next, we describe the JWST data of the Cosmic Cliffs field and highlight recent work done to identify YSOs with JWST data (\S\ref{sec:jwst_data_bg}).}

\subsection{NGC 3324 and Gum 31} \label{sec:target_area}

\begin{figure*}[t]
    \centering
    \includegraphics[width=\textwidth]{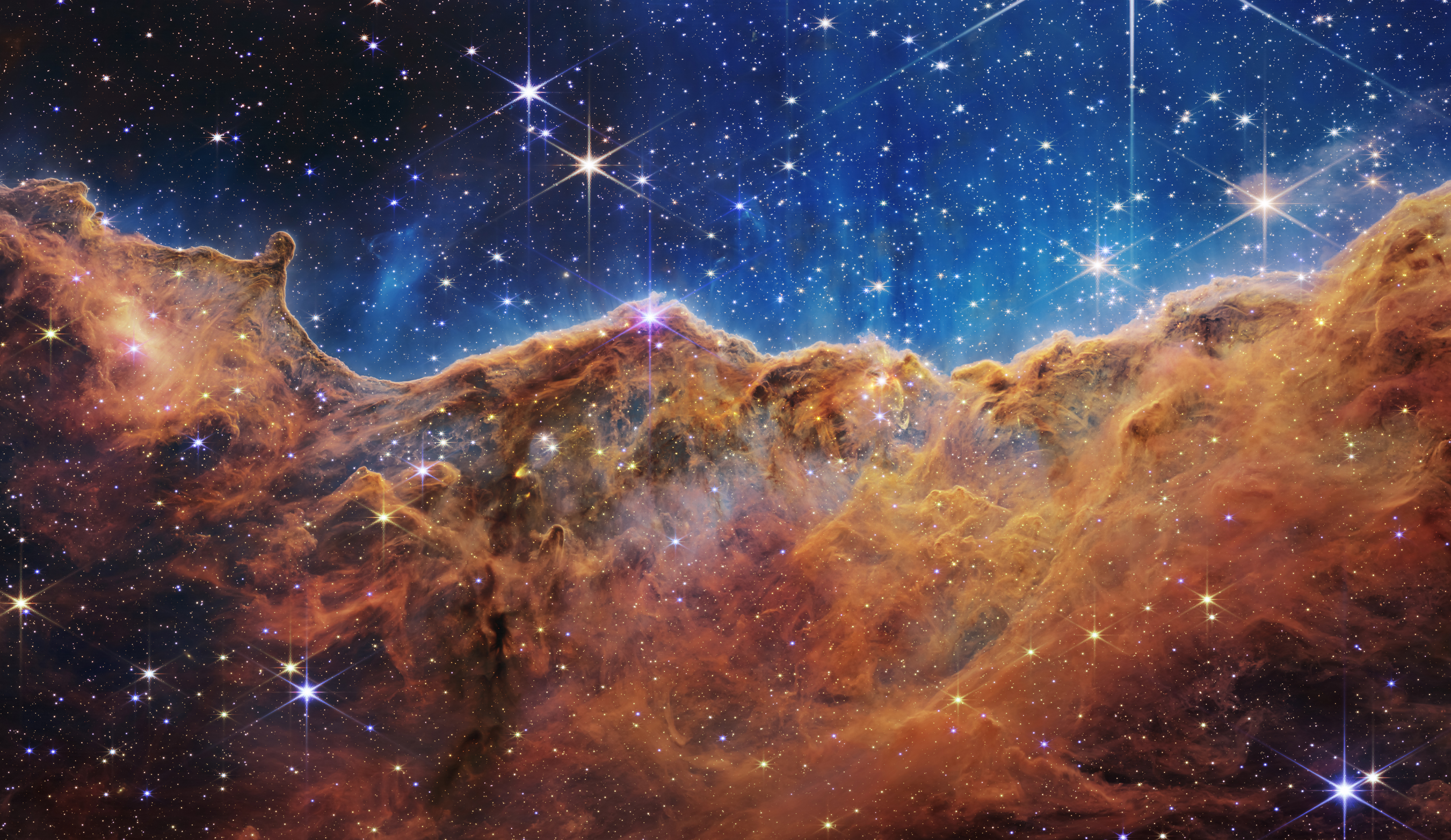}
    \caption{The region of NGC 3324 imaged by JWST. Stiff peaks along the Gum 31 H-II region are formed by stellar winds originating from the massive stars in NGC 3324. This color image is made from six NIRCam filters. Red: F444W, Orange: F335M, Yellow: F470N, Green: F200W, Cyan: F187N, Blue: F090W. Image credit: NASA, ESA, CSA, and STScI.}
    \label{fig:jwst_stsci_im}
\end{figure*}

NGC 3324 is a young stellar cluster located on the North-Western corner of the Carina Nebula Complex, inside the Gum 31 HII region. The diameter of Gum~31 is $\sim 15'$ \citep{Cappa2008}, while NGC~3324 is $11'$ across \citep{Bisht2021}. Recent studies of the young cluster have placed it at a distance of 2.35 kpc \citep{Goppl2022}, while the HII region is attached via filamentary structures to the larger Carina Nebula Complex \citep{Preibisch2012,Roccatagliata2013}. NGC~3324 includes at least three massive stars, one of which (HD 92207, spectral type A0Ia) has a very strong stellar wind \citep{Kruditzki1999}. This cluster excites the Gum 31 nebula, and causes sharp features to form along the edges of the bubble, as seen in the JWST/NIRCam image shown in Figure~\ref{fig:jwst_stsci_im}.

Previously, the Gum 31 region was examined with Spitzer, WISE, and Herschel data by \citet{Ohlendorf2013} to search for YSOs. They found a total of 661 candidate YSOs within the WISE data, as well as 91 protostar candidates within the Herschel data. Of these, 17 sources were fit to SED models, and a completeness limit of $\sim1~M_\odot$ was determined. Based on these results, they extrapolated the initial mass function to determine a total population of 5000 YSOs with masses greater than $\sim 0.1$ $M_\odot$ within the Gum 31 nebula. From spatial distributions of the candidate YSOs, they concluded that both collect-and-collapse and radiative triggering are the prevailing avenues of star formation actively ongoing within Gum 31. 

Further Herschel data of the greater Carina Nebula, which connects to Gum 31 and is the larger portion of the overall complex, were analyzed by \citet{Gaczkowski2013}. They found 144 protostars with detections in the 70 $\mu m$ band, and estimated the total protostellar population to have 2880 objects, based on the Kroupa initial mass function.  From this estimate, they derived a star formation rate within the full Carina Nebula Complex of $\sim 0.017$ $M_\odot/yr$. A spatial analysis of all 642 detected sources revealed that the Class 0 and Class I sources are located just within the cloud edges, while the more-evolved YSOs identified in previous surveys are located beyond the cloud edges. 


Gum 31 was examined in the x-ray by \citet{Preibisch2014xray}, who found 679 x-ray emitting objects, 500 of which were matched to objects also detectable in the IR. Although x-rays trace YSOs, they trace other sources as well, such as fore- and background field stars and extragalactic sources. As such, \citet{Preibisch2014xray} determined that the contamination by such sources within their sample is likely to be 25-30\%. They found, however, that those objects with mid-IR counterparts, found using VISTA, make up approximately 75\% of the catalog, and proposed that these objects are true YSOs and not contaminants. 

\citet{Preibisch2014xray} also analyzed the x-ray luminosity function and determined there should be approximately 4000 YSOs of mass greater than $\sim 0.1$ $M_\odot$ within the associated emission nebula of Gum 31, which falls within the estimate of \citet{Ohlendorf2013}. They found that 30\% of YSOs are concentrated in two areas, one of which is NGC 3324, and that the remaining 70\% are spread homogeneously across the full bubble. By comparing the sizes of Gum 31 and the Cosmic Cliffs and using the above distribution, their results suggest approximately 630 YSOs should exist within the Cosmic Cliffs field alone, including many Class III objects. 

\subsection{ML Tools}\label{sec:ml_tool_bg}
Within the last several years, ML has emerged as a useful tool within star formation studies for classifying YSOs. 
Several ML approaches have been explored, including Gradient Boosting \citep{Miettinen2018}, neural networks \citep{Miettinen2018, Cornu2021, Chiu2021}, and RF models \citep{Kuhn2021}.  Each of these ML approaches has its strengths, and each has successfully separated YSOs from the contaminating objects in the fields studied such as stars, galaxies, knots of PAH emission, and Active Galactic Nuclei (AGN) to high accuracy. 

There are many different ML algorithms to choose from, and for context we summarize briefly here those mentioned in this paper. There are two main branches of ML: classical and neural networks. Classical approaches include Random Forest (RF) algorithms \citep{ho1995random} that make use of a set of decision trees to determine either a class (based on the class returned most often) or the value of a requested parameter (via an average). Gradient Boosting \citep{friedman2001greedy} is another classical approach that is similar to RF, with each successive tree learning from the errors of its predecessor. Although this approach may be favorable, it can also lead to over-fitting and can be unable to generalize to new data.  Neural networks also come in a variety of forms. The typical artificial neural network (ANN) is made to mimic the brain, with ``neurons'' connected to nodes that can have connections of different weight. The more nodes and more layers of connections an ANN has, the more specialized it can be.

Recently, \citet{Kuhn2021} classified YSOs using data from the MYStIX catalog \citep{Povich2013} across the entire Galactic Plane. One subsection of this dataset is the GLIMPSE Vela-Carina survey \citep{Majewski2007}, which imaged the entire Carina Nebula Complex, including the Cosmic Cliffs. They utilized an RF model trained on an imbalanced\footnote{{The numbers of objects within each defined class are highly dissimilar.}} mix of YSOs and contaminant (non-YSO) field objects, where the number of YSOs accounted for less than 25\% of the full training set. {They used a standard ML metric called ``Area Under the Curve" (AUC), where the curve in question is the Receiver-Operator Curve (ROC), to determine the best fits. A ROC curve shows a binary classifier's ability to distinguish true from false classifications, as the discrimination threshold changes.} 
As RF models output the probabilities of objects being in the positive class, \citet{Kuhn2021} defined an object as being a YSO if it had at least a 50\% probability of being so determined by their RF model.
After YSOs were identified as such, \citet{Kuhn2021} further used cuts based on spectral index to determine their respective stages within pre-main-sequence evolution, thereby labelling them as either Class I, Flat-Spectrum, or Class II YSOs. They published their catalog as the Spitzer-IRAC Catalog for YSOs, henceforth referred to as SPICY.

Though powerful, ML classification algorithms can have pitfalls.  First, imbalances in the numbers of objects in different classes can make it difficult to train an ML algorithm to classify objects equally well.
As YSOs are much less numerous than regular field stars, such imbalances are a relevant issue when identifying YSOs via ML. RF models, such as that used by \citet{Kuhn2021}, however, are better at handling imbalanced datasets, by individually classifying each object based off of a training set \citep{Breiman2001}. 
Second, most ML algorithms cannot handle missing data. \citet{Kuhn2021} worked around this issue by using copulas, i.e., functions that connect the probability distributions of different features to each other.  When using copulas, the joint probability distributions of the data are broken down into their base components, and the copulas combine these probabilities together \citep{Nelsen2007}. Using copulas to fill missing data, however, assumes that data are missing because an object is not detected in the field observed by the particular filter. 

\citet{Chiu2021} used a different approach to missing data that assumes objects in question are 
indeed present but not seen as point sources.  For example,
a YSO may be heavily obscured in certain filters, or its SED may drop below the sensitivity limit of the instrument.  The solution provided by \citet{Chiu2021} was to fill the data in missing bands with 1\% of the smallest flux obtained in that band as a reasonable estimate for the thermal noise of the detector. This method hence accounts for the clarity of an object at different wavelengths, which is important for the determination of class.

An alternative solution to the issue of missing data is provided by the Probabilistic Random Forest (PRF) method released by \citet{Reis2019}, which has successfully been applied to high-mass YSO identification in Local Group galaxies \citep[e.g.,][]{Kinson2021,Kinson2022}. The PRF method uses both the values and errors of each filter to create probability distributions for each data point, where the expectation value is the data point's flux, and the standard deviation is the error on this flux, assuming a Gaussian distribution. An RF-like algorithm is hence trained, and when an object is sent through the network, it is no longer sent along one branch of the tree. Instead, at every decision node, the probability of the object being on either side of the node is propagated, with probability determined by the Gaussian distribution. For a full prescription, see \citet{Reis2019}. This method has the benefit of not assuming what the missing data may be while still accounting for them by passing any node that relies upon the data with equal probability to either side. 

The very dusty field of the Cosmic Cliffs means there are many objects within it that are not visible at all wavelengths. In particular, YSOs are often deeply embedded in dust, and so we cannot ignore objects that are missing data, particularly at short wavelengths. Furthermore, the high sensitivity of JWST means that many brighter objects are saturated, and hence do not have photometric data despite being `observable.'
For this paper, we choose to evaluate the Probabilistic Random Forest method for our classification, applying this method to the new Cosmic Cliffs data from JWST. We build off previous work to create a larger catalog of YSOs within the Cosmic Cliffs region, which is accurately mapped to JWST photometry. In so doing, we demonstrate the usefulness (and limitations) of the PRF methodology for analyzing JWST data, and from there quantify the improvements in detecting YSOs now possible with JWST.

\subsection{JWST Data}\label{sec:jwst_data_bg}

On July 11th, 2022, the first observations from JWST were released \citep{Pontoppidan2022}. These data included observations of four different astrophysical objects, one of which was part of the NGC 3324 star-forming region in the larger Carina Nebula Complex. JWST imaged this region with two instruments: the Near Infra-Red Camera (NIRCam) to observe its dust and stars and look for emission lines of H$_2$, Poly-Aromatic Hydrocarbons (PAHs), and Pa-$\alpha$; and the Mid Infra-Red Instrument (MIRI) which was able to probe for objects hidden within the dust that may have been rendered undetectable at shorter wavelengths.


\begin{deluxetable}{ccl}
\tablewidth{\textwidth}
\tablecaption{JWST filters used by ERO to image NGC 3324, their exposure times, and their uses as described in \citet{Pontoppidan2022}.
    \label{tab:filters}}
\tablehead{\colhead{Filter}& \colhead{$t_{exp} (s) $} & \colhead{Use}}
    \startdata
        F090W & 25768.32 & dust and background stellar field\\
        F187N &46382.88& ionized gas via the bright Pa-$\alpha$\\
        F200W & 25768.32&dust and background stellar field\\
        F335M & 6442.08& 3.3 $\mu m$ PAH emission\\
        F444W & 6442.08 &dust scattering from large grains\\
        F470N & 11595.72 & H$_2$ from embedded jets/outflows\\\hline
        F770W & 6771.08 &PAH emission\\
        F1130W & 6771.08 &PAH emission\\
        F1280W & 6993.12 &12.81 $\mu m$ [Ne II] line emission\\
        F1800W & 5994.08& cool dust and proplyds\\
    \enddata
        \tablecomments{ The first block lists NIRCam filters and the second lists MIRI filters.}
\end{deluxetable}

The Early Release Observations \citep[ERO,][]{Pontoppidan2022} of NGC 3324 focused on the Cosmic Cliffs, a $\sim$7\farcm4 $\times$ 4\farcm4 area with NIRCam, and a $\sim$6\farcm4 $\times$ 2\farcm2 area within the NIRCam field with MIRI, all located on the edge of the Gum 31 bubble. The data were collected in six NIRCam bands and four MIRI bands; see Table~\ref{tab:filters} for details on the bands used, exposure times, and intended uses. 
The exposure times varied for each filter, and the \texttt{FULLBOX} 10-point dither pattern was used for NIRCam imaging and 8-point dither for MIRI imaging \citep{Pontoppidan2022}. We will henceforth focus on the NIRCam data for this paper as, due to its much wider field of view, NIRCam will be used more extensively of the two, and indeed often on its own, to probe YSO populations.

Reduced FITS images, as well as a source catalog generated for each filter by the JWST pipeline, were made publicly available through the Mikulski Archive for Space Telescopes (MAST)\footnote{ {\dataset[DOI:10.17909/67ft-nb86]{\doi{10.17909/67ft-nb86}} \citep{JWST-ERO-DOI2022}}}.
For our use, {we accessed the available FITS images and made our own source catalogs from them, as those provided by MAST were deemed inadequate for our intended purpose.}
These FITS images have been reprocessed several times since July 2022 as better JWST flux calibrations became available. For this paper, the data products for NGC 3324 were last accessed on {June 13th, 2023.}

The JWST data of the Cosmic Cliffs have already been probed to understand the capabilities of JWST to detect jets and outflows. Previously, \citet{Reiter2022} looked at data from the narrowband 1.87 $\mu$m filter and the difference between data from the narrowband 4.7 $\mu$m and wideband 4.44 $\mu$m filters from JWST. In combination with archival Hubble data, they used these datasets to identify 31 outflows within the Cosmic Cliffs field of view, including seven Herbig-Haro objects only visible in the infrared (IR). Along with their identifications of outflows, they provided a list of 21 outflow driving source candidates, i.e., IR-excess sources located along the estimated arcs of travel determined by tracing the observed outflows back in time. As not all outflows were visible in the earlier Hubble data, they could not determine proper motions for the entire sample, and as a result straight-line estimation was used when appropriate. 
\citet{Reiter2022} also checked to see if any of their identified YSO candidates had been previously identified with Spitzer data via comparison with SPICY, and found matches to six of the 21 possible outflow driving source candidates.

{More recently, an analysis of JWST data of the nearby ($\sim 490$ kpc) star-forming dwarf galaxy NGC 6822 has illustrated the importance of using JWST data for YSO identification. \citet{Lenkic2023} found that YSOs selected using Spitzer-based criteria could be matched to as many as six different objects in JWST images that had been previously blended together in lower-resolution Spitzer data. They additionally were able to identify nearly three times as many YSOs as in previous surveys, and clarified that several ``YSOs" identified in previous surveys were instead either reddened stars or galaxies, further emphasizing the improvements of JWST over Spitzer.}

\section{Methodology}\label{sec:Method}
In this section, we verify the PRF model as a reasonable model for classifying YSOs (\S\ref{sec:PRF_test}), describe the creation of the catalog and the training/validation sets of objects in the Cosmic Cliffs field (\S\ref{sec:catcreate}), and finally describe how well the PRF method works for this study (\S\ref{sec:PRF_apply}). 

\subsection{Testing the Validity of the PRF method} \label{sec:PRF_test}

{We compare here the results of the regular RF to the PRF on a test dataset.}
\citet{Reis2019} provided a comparison with the regular RF that shows that when all labels are correct, the PRF and RF perform at the same accuracy. When purposefully introducing incorrect target labels, however, they found that the PRF greatly outperforms the regular RF.  As an additional test, we demonstrate the relative performance of a PRF vs. a regular RF approach, using both copulas and thermal noise to fill the missing data.  For this test, we use data from the Cores to Disks \citep[c2d,][]{Evansc2d} survey, which contains Spitzer data with both completely filled and missing values. We first use 10 000 objects with all bands available, then 9000 objects with all bands available and an additional 1000 (randomly chosen) objects with data missing in at least one band to obtain a case where 90\% of data are filled. Similarly, we also obtain data-sets where 80\%, 70\%, 60\%, and 50\% of data are filled. In all cases, the data are real observations, and no data are artificially removed. We use YSOs as our positive class and all others as contaminants, where YSOs make up approximately a third of the sample. 

{We focus on four metrics that can be used to assess the performance of an ML model:} accuracy, recall, precision, and F1-Score. Each of these requires some combination of the numbers of True Positives (TP), False Positives (FP), True Negatives (TN), and False Negatives (FN). For our purposes, TP is the number of objects correctly classified as YSOs, TN is the number of objects correctly classified as contaminants, FP is the number of contaminant objects incorrectly classified as YSOs, and FN is the number of YSOs incorrectly classified as contaminants. Accuracy, $A=(TP+TN)/(TP+FP+FN+TN)$, is a measure of the total number of correct identifications but can be easily made misleadingly high as a result of having a relatively populous negative class.  As such, we do not use it here. The F1-score, F1 $= 2 R\times P/(R+P)$, however, is a metric defined as the harmonic balance between recall $R = TP/(TP+FN)$ and precision $P = TP/(TP+FP)$. We use F1-score as our metric of choice because we are aiming to configure a network with low contamination by much more numerous contaminants (high precision) while still maintaining a high recovery of YSOs (high recall). 

\begin{figure}
    \centering
    \includegraphics[width=\columnwidth]{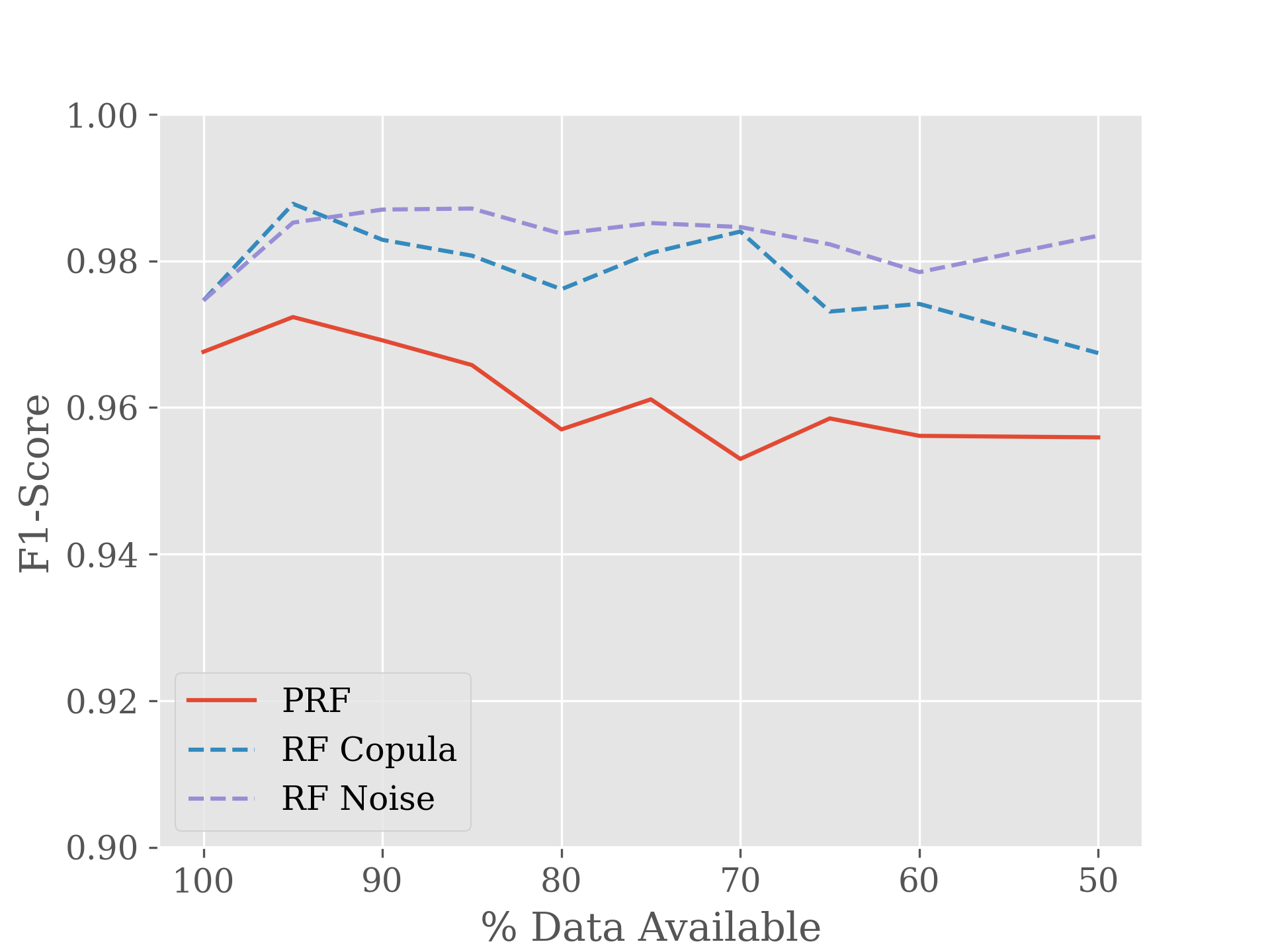}
    \caption{A comparison of the F1-Scores for the validation set of the PRF (solid red line), an RF filled via copulas (blue dashed line), and an RF filled with thermal noise (purple dashed line) as a function of the amount of missing data.}
    \label{fig:RFvsPRF}
\end{figure}

Figure~\ref{fig:RFvsPRF} shows the F1-Score performances for the RF and PRF models when the amount of data available is decreased.  Two RF methods are evaluated, one with copulas and the other with thermal noise used to fill missing data.
We used the \texttt{Python} package \texttt{copulas} for our calculations. 
{A slight decrease in F1-Score is seen with the increase in objects that are missing flux measurements.}  Nevertheless, we find that all three cases perform well ($>95\%$) and within a few percent of each other, though the PRF F1-Scores are slightly but systematically lower than the RF scores. In general, filling the data with noise or using copulas performs equally well. 

{Despite the high proficiency of using copulas or thermal noise, these methods come with their own caveats. The use of copulas assumes that the data were not imaged and that the environment that generated the copula is reflective of that in which the point source in question is missing data, which is not necessarily the case for point sources missing JWST data. Similarly, filling the data points with thermal noise assumes that the object was not detectable because it was too weak at a particular wavelength, and does not take into account that an object may be saturated, or that the source was not imaged at all. Both cases rely on certain assumptions, which may or may not be true for any given point source. The PRF, however, does not make any assumptions about why data are missing, and instead works around them. Although this method is much simpler than the above methods for filling missing data, it still performs quite well in the above test. Indeed, this paper serves as a proof of concept for applying the PRF method to JWST data.} 

{For this paper, we choose not to fill the data with either noise or copulas when comparing to the regular RF model, as in some cases the objects are not measured in the band due to over-saturation, and others because they are not visible due to intervening dust or not detected at that wavelength. In this way, we proceed with one algorithm which we allow to make its own assumptions about the missing data (PRF) and one algorithm which makes no assumptions about the missing data and instead ignores those objects (RF).}

\subsection{Catalog Creation} \label{sec:catcreate}
We used pipeline-generated JWST images of the Cosmic Cliffs fields that were made publicly available as FITS files by MAST.  {The specific data analyzed in this work can be accessed via \dataset[DOI: 10.17909/bsbn-3m50]{https://doi.org/10.17909/bsbn-3m50}.} 
We did not use the photometric data also provided by MAST for two reasons: (1) a catalog of all identified sources was not accessible, and, more importantly, (2) many objects were found to have photometric data in only a subset of the bands for which they are visible. For example, an object's fluxes could be provided for the F090W and F200W bands but not the F444W band, despite the object being clearly visible in the F444W image.
Thus, we applied the photometric software DAOPHOT \citep[][version as of May 30th, 2023, accessed via private communication]{Stetson1987,Stetson2011} to the NIRCam data, keeping only objects that were seen in at least three bands that had measurement errors less than 0.2 mags.  With these criteria, we retrieved a total of \totobjs{} individual sources in the Cosmic Cliffs NIRCam field. The number of objects in each NIRCam band is listed in Table~\ref{tab:photnum}.


The DAOPHOT program obtains photometry by first identifying all point and extended sources within the field, and running synthetic aperture photometry on them. From this photometry, a local sky brightness and magnitude is computed for each object.  {These local values are used to compute a background around each source, thereby minimizing the effects of variable background emission from the nebula.} Finally, an empirical point-spread function (PSF) is determined for each frame and used to compute PSF photometry for each object. Importantly, our PSF is defined from 200 user-identified sources in each band. These sources must (a) not include any bad or saturated pixels; (b) be free from the bright dusty fields and hence from any stray emission which could confuse the PSF algorithm; and (c) must be within 2-sigma of the mean PSF from all identified PSF stars. Each source is fitted to the PSF using least-squares profile fits. 
 {In all cases, the reported error on the calculated magnitude is a compromise between the observed pixel-to-pixel root mean square, and the uncertainty predicted from photon statistics and read noise, whichever of the two is larger.}


\begin{deluxetable}{cc}
\tablewidth{\textwidth}
\tablecaption{The number of objects with photometry in each band.
    \label{tab:photnum}}
\tablehead{\colhead{Band}&\colhead{Number}}
\startdata
      F090W & 19~061 \\
       {F187N} 
      &  {18~329}\\
      F200W & 19~490 \\
      F335M & 19~458 \\
      F444W & 19~486 \\
      F470N & 19~427 \\
\enddata
    
\end{deluxetable}

At this point, we excluded some of the F187N data from our analysis due to a positional misalignment relative to other images at approximately x = 6000 pixels. {The photometry calculated for the data around this misalignment ($\pm 5$ pixels) was disregarded and set to NaN.}
When we attempted to use those data regardless, spurious YSO detections were obtained within 500 pixels to either side of the misalignment.  These no longer occurred when {these data were} removed from the ensemble.

{Next, we chose 253 Spitzer-detected objects within the Cosmic Cliffs field that could be securely matched to a bright and relatively singular counterpart in the JWST data. Appropriate sources were hand-chosen based on their magnitude, where only JWST objects with bright fluxes that were coincident with Spitzer sources (i.e., within the latter's FWHMs) were chosen.  We removed any sources that were blended or which we were unable to identify their JWST counterpart confidently.} We next used the \texttt{Astropy} \citep{Astropy} \texttt{match\_coordinates\_sky\_} task to match the Spitzer detections from the GLIMPSE catalog of the region \citep{Majewski2007}\footnote{\doi{10.26131/IRSA213}}, available SPICY targets \citep{Kuhn2021}{, and appropriate YSOs from \citet{Ohlendorf2013}, resulting in a total of \ysoav{} YSOs: 18 from SPICY, 11 from \citet{Ohlendorf2013}, with six objects overlapping. The locations of these objects were then updated to match their hand-selected JWST counterparts. This process resulted in \totav{} objects detected by Spitzer being matched to JWST sources, of which \contav{} are contaminant objects, and \ysoav{} are YSOs}.

Once the catalog was created, we utilized data augmentation to 
create a simulated dataset for training purposes.  For this approach, we first removed any identified sources with missing data,
leaving us with \ysowomiss{} YSOs and \contwomiss{} contaminant objects from which to build an augmented dataset.  

{We tested two methods of data augmentation to expand our training set. In both cases, we withheld 10\% of the data to confirm the generalization of our network, and so the augmented dataset was made from the data of \ysotrain{} YSOs and \conttrain{} contaminants.}
{The first method, which we will refer to as the Error Variation Technique (EVT), creates more sources to train the model by shifting the SEDs of objects of known class within the errors on their photometric points. This method was first used in \citet{Cuperlovic2021}, and is similar to the method from \citet{Shy2022}.} {With EVT, we determined the mean error in the flux for each band. This error was then multiplied by a random real value between 0 and 3.}
We then randomly sampled and varied the data from objects in each class by adding or subtracting this error, efficiently producing new sources that were still, within {3 sigma} error, effectively the same as the original sources and with which we could thus securely apply classifications.
{The second method, the Synthetic Minority Over-sampling Technique \citep[SMOTE;][]{SMOTE2002}, creates more data by interpolating between features of the same class. We utilize the \texttt{imblearn SMOTE} python library to implement SMOTE. Both strategies allow us to over-sample the smaller YSO class while still producing unique data 
with which the algorithm can be trained. }

Finally, we split our data into training and validation sets. The training set is made of 20 000 total objects obtained via the aforementioned data augmentation for each class. The validation set uses all \totav{} objects with classifications (\ysoav{} YSOs and \contav{} contaminant objects). We maintained this imbalanced set to ensure that the metrics we obtained were representative and that we did not produce misleading metrics resulting from under-sampling the contaminant class or over-sampling the YSO class. In addition to having data that are, by design, similar to the augmented data, this validation set also has \tottestprf{} objects which are completely unknown to the augmented data. 

\begin{figure}
    \centering
    \includegraphics[width=\columnwidth]{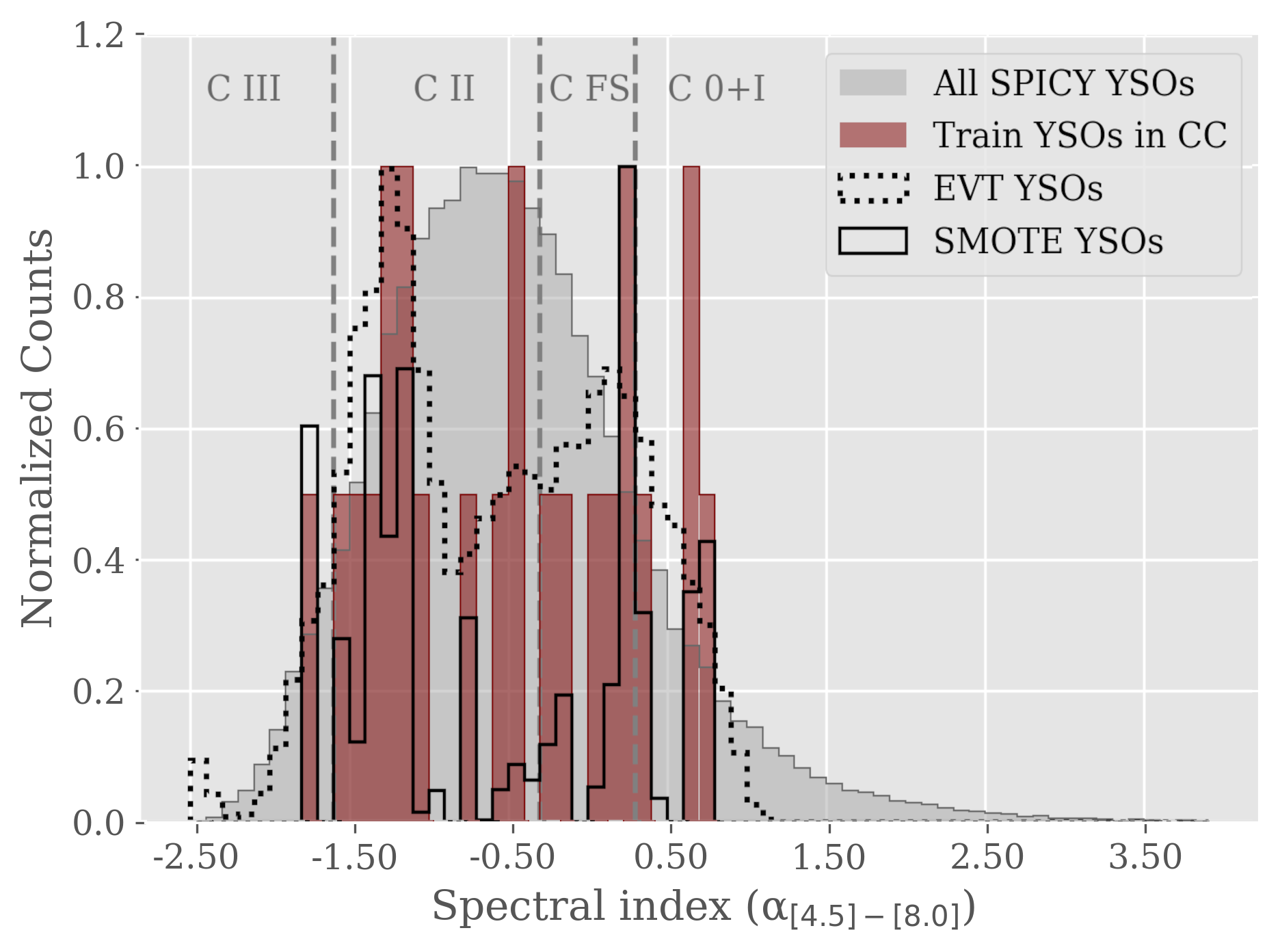}
    \caption{{The spectral indices for the YSOs which form the basis for our training set (maroon), an example of how the EVT method (dotted black line) or SMOTE method (solid black line) of data augmentation can cause variation in spectral index, and the full range of spectral indices throughout the entire SPICY catalog (gray), normalized for comparison. Gray dashed lines indicate the typical cuts in spectral index applied to separate classes \citep[e.g.][]{Dunham2015,Kuhn2021}. The associated class labels are also marked above their respective regions (Class II $\equiv$ ``C II").}}
    \label{fig:spectral_index_comp}
\end{figure}

{The set of objects used to build our training set is relatively small, and as such the types of objects classified as YSOs will be limited by the diversity of this set in parameter space.}
{To ensure the set of objects that comprises our training set is diverse enough to function as a general set, we analyze the spectral index distribution of the YSOs from Spitzer data, specifically the index from the IRAC 2 to IRAC 4 bands, or 
\begin{equation}
    \alpha_{[4.5]-[8.0]} = 1.64([4.5] - [8.0]) - 2.82
    \label{eq:alph}
\end{equation}
We compare this set of spectral indices to that of the full SPICY catalog over the entire GLIMPSE dataset. Figure~\ref{fig:spectral_index_comp} shows a histogram of the spectral indices of the Cosmic Cliffs YSOs identified by SPICY against those of the entire SPICY sample.
Missing in the data used to build the training set are only YSOs with extreme spectral index, primarily the Class 0 or more-embedded Class I YSOs that are difficult to detect in the near-infrared in the first place and the very stellar-like Class III YSOs. }  In particular, the SPICY catalog lists 14 Class II YSOs ($\sim 54\%$), five Class I YSOs ($\sim 19\%$), six flat-spectrum YSOs ($\sim 23\%$), and one YSO of uncertain type ($\sim 4\%$) within the Cosmic Cliffs.

{Figure~\ref{fig:spectral_index_comp} also shows how each data augmentation method can cause variation in the spectral indices and hence the objects we can identify. In each case, the distribution is more uniform and expands slightly beyond the bounds of the original set, with the EVT method showing greater variation than the SMOTE method. 
Hence, we expect that the PRF algorithm will successfully identify a wide range of YSOs in the Cosmic Cliffs, including less embedded Class I YSOs and Class II YSOs.}
{Indeed, it is beneficial to utilize YSOs from the same region as a training set for this initial study.  At similar distances and experiencing similar extinctions, they arguably exhibit signatures that are most similar to those of the YSOs in the region we aim to identify.  }

Though Figure~\ref{fig:spectral_index_comp} demonstrates that the training set is composed of YSOs that cover a wide range of pre-main-sequence evolution, as evidenced by the spectral index, the relatively small size of this training set may yet mean the full range of YSO colors at shorter NIR wavelengths is not similarly unbiased. For example, the YSOs in our training set may not represent the range of NIR colors that occur from disks viewed over a wide range of inclinations or those transition disks whose inner regions have been cleared out. For the present study, we aim to evaluate the performance of the PRF method given JWST and Spitzer data in hand, though a larger training set would indeed provide a fuller basis for identifying a wider range of YSOs in the Cosmic Cliffs field. Such training sets will become available in future when more JWST images of star-forming regions previously observed by Spitzer become available for public analysis. Until then, we proceed with the understanding that the limited training set used here likely results in a set of identified YSOs that comprise a lower limit to the true number within the Cosmic Cliffs.

\subsection{Applying the Probabilistic Random Forest method} \label{sec:PRF_apply}
{We analyzed data from the ERO of the Cosmic Cliffs region in six NIRCam filters, those corresponding to 0.90 $\mu$m (F090W), { 1.87 $\mu$m (F187N),} 2.00 $\mu$m (F200W), 3.35 $\mu$m (F335M), 4.44 $\mu$m (F444W), and 4.70 $\mu$m (F470N).  {In these bands, JWST was able to achieve} sensitivity limits of 28.6 mag, { 25.0 mag,} 26.2 mag, 22.1 mag, 22.8 mag, and 22.3 mag, respectively. Conversely, the  {brightness} of objects {was} able to be retrieved up to magnitudes of 12.9 mag,  {10.0 mag, }10.5 mag, 8.2 mag, 7.7 mag, and 6.7 mag, respectively, where objects beyond these limits have too many saturated pixels for accurate photometry. 
Indeed, the saturation limits of the longer wavelength range are counterproductive to the identification of YSOs, as that range is where the YSOs will be at their brightest in the NIRCam data. 
 {As a result, YSO numbers may be slightly underestimated.}
Fortunately, this saturation limit is only relevant for just 60 of \totobjs{} total objects within the Cosmic Cliffs field, and so should not greatly impact the results. Hence, previous IR studies with lower sensitivity can be still complementary to those  {performed} with JWST.}

{To classify the full set, we utilize the aforementioned training and validation sets in the following manner.}
The PRF requires three input parameters for supervised classification: the input data, errors on the input data, and either the targets for the input data or the probability of the object having the positive or negative label.  After some testing, we determined that using the targets rather than the probabilities was more appropriate. We use only colors as the input data to eliminate the bias of only choosing objects of high flux due to using Spitzer-identified objects as the basis for our training set. We hence employ every combination of filters possible, e.g., F090W-F187N, F090W-F200W, F090W-F335M, etc.

We tested next the two different data augmentation techniques, EVT and SMOTE. To determine which to use going forward, we {first} compared the contamination rates of our validation set for each. We found that classification using SMOTE had a higher contamination rate than using EVT ( {10}/185 vs.  {1}/185), even when we used the full set of \ysowomiss{} YSOs and \contwomiss{} contaminants to build the SMOTE set. {Second, when} extrapolating to the full dataset of \totobjs{} objects utilizing the SMOTE-augmented data for training, the PRF method found all of the YSOs identified with the EVT-augmented data. The number of additional YSOs found when using SMOTE{, however,} is nearly triple its contamination rate ($\sim14.5\%$ compared to $\sim5.4\%$). {For these reasons, we opt to use EVT-augmented data and not SMOTE-augmented data for our classifications.} 

{Although EVT works better here, it may not strictly do so in all possible situations - we have not performed a thorough comparison of the two approaches.  Nevertheless, SMOTE can suffer from issues where synthetic data points deviate from real data points in complex parameter space \citep[see, e.g.,][]{Bellinger2016}.  This behavior can be caused by generating points within the convex-hull, where SMOTE generates synthetic data points along lines connecting all points in the parameter space rather than following the curve of such points as it varies across parameter space.  EVT, however, circumvents this issue by only filling in synthetic data around real data, and thereby follows the trend of points across parameter space.  As a result, EVT produces a population of synthetic data that covers the parameter space without straying from the trend of real data, albeit unevenly.  The EVT method is thus more likely to minimize the false positive rate.}


 {In either case,} the number of identified objects varies from run to run, despite the F1-Scores remaining the same. 
Hence, we take the classifications for all objects over 100 runs of the model and take the probability of a given object being a YSO as the average probability that it has been so identified. The error on the assigned probabilities is $\sim1\%$. With all bands included in the ensemble, \totysos{} objects are identified in the Cosmic Cliffs as having a $>50\%$ probability on average of being a YSO  {using the EVT method for data augmentation}. 

\begin{figure}
    \centering
    \includegraphics[width=\columnwidth]{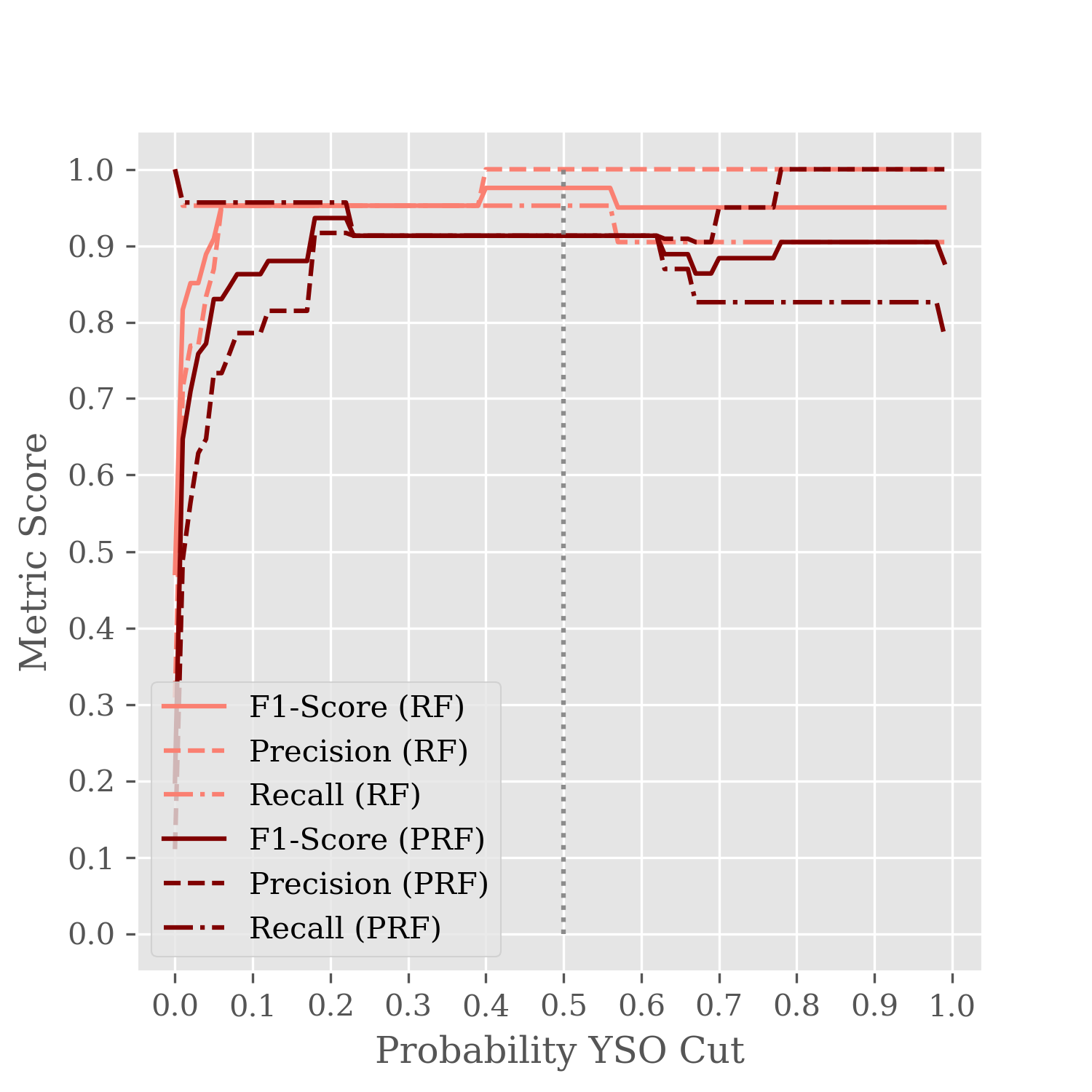}
    \caption{The trends of various metrics with a variation of the cut in probability that determines if a given object is a YSO. There are no YSOs with exactly 100\% probability. }
    \label{fig:prob_yso_metric}
\end{figure}

Figure \ref{fig:prob_yso_metric} shows various metrics for the validation set of the YSO classification as a function of the threshold on the probability of a given object being a YSO. The higher the threshold, the fewer objects are classified as YSOs, leading to a general decrease in recall and increase in precision, though the F1-Score remains relatively constant. 
 {The RF models tend to have higher metrics as a result of the augmented data used as training being based on all but \ysotest{} of the YSOs and \conttest{} of the contaminants that the RF is testing, while the PRF has an additional \ysoadtest{} YSOs and \contadtest{} contaminants not included in the augmented data. When the objects with missing data are removed, however, the PRF performs equally well. As we wish to evaluate the objects with missing data, we choose to evaluate our metrics on the full dataset.}
There are no YSOs with exactly 100\% probability with the RF, leading to the sharp decrease in all metrics at that extreme.

\begin{figure}
    \centering
    \includegraphics[width=\columnwidth]{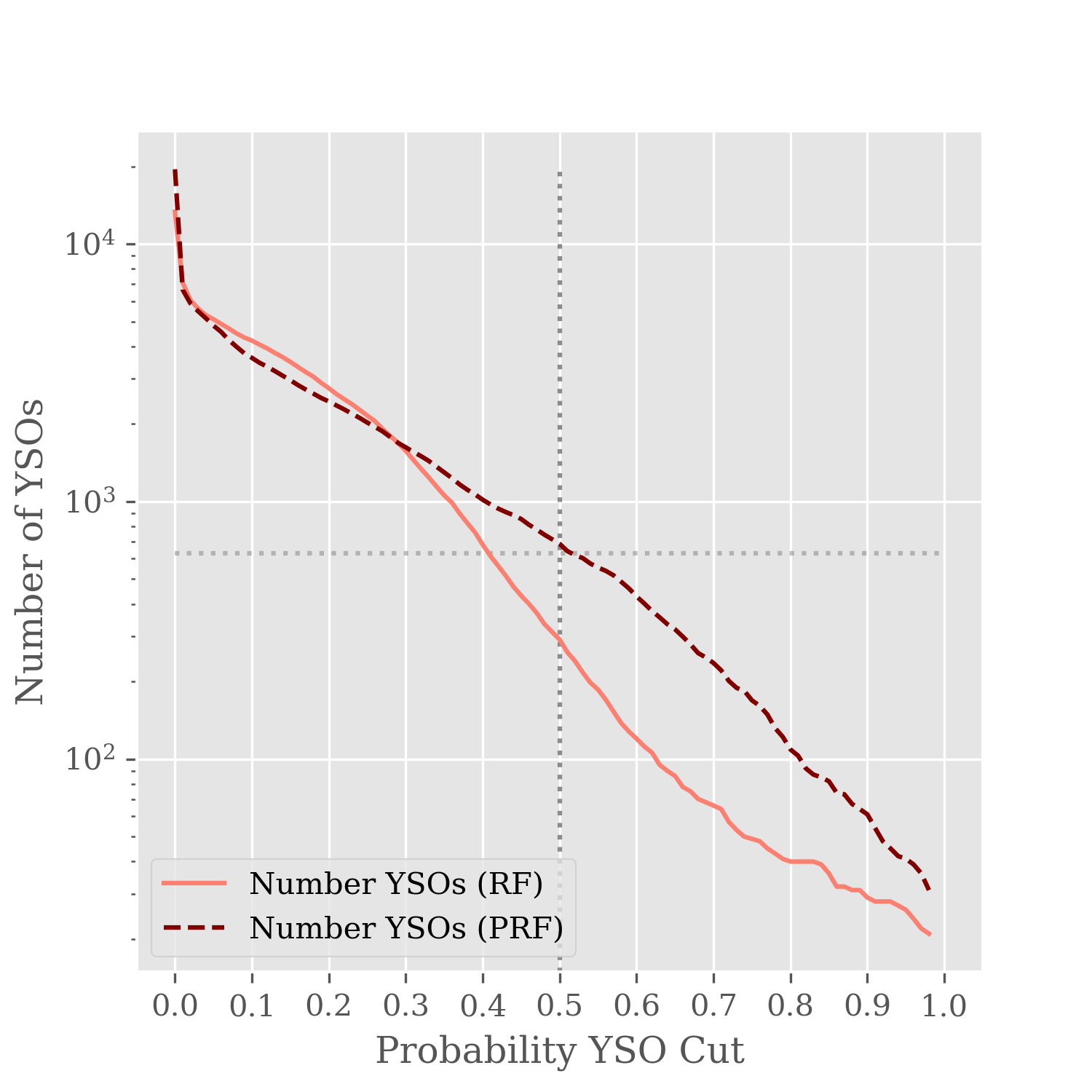}
    \caption{The trend of number of YSOs with  the cut in probability that determines if a given object is a YSO. There are no cYSOs with exactly 100\% probability. The gray horizontal line marks the upper limit of YSOs we expect within the region from previous works, and the gray vertical line marks the 50\% threshold.}
    \label{fig:prob_yso_num}
\end{figure}

{Figure \ref{fig:prob_yso_num} shows the numbers of YSOs that are identified via the RF and PRF when the minimum threshold probability of being a YSO is increased.  The threshold ultimately selected is arbitrary, and so we set it to 50\% to require that the probability of the object being a YSO is greater than the probability it is not. The 50\% threshold is standard for ML purposes, and is further justified in our case because our analysis is based on YSOs identified previously in the Cosmic Cliffs by \citet{Kuhn2021} with the same threshold. 
All PRF algorithms output the probability of an object being a YSO or contaminant.  Again, we eliminate the bias of the random seed on the classification of objects by taking the average probability over 100 runs.}

\begin{figure}
    \centering
    \includegraphics[width=\columnwidth]{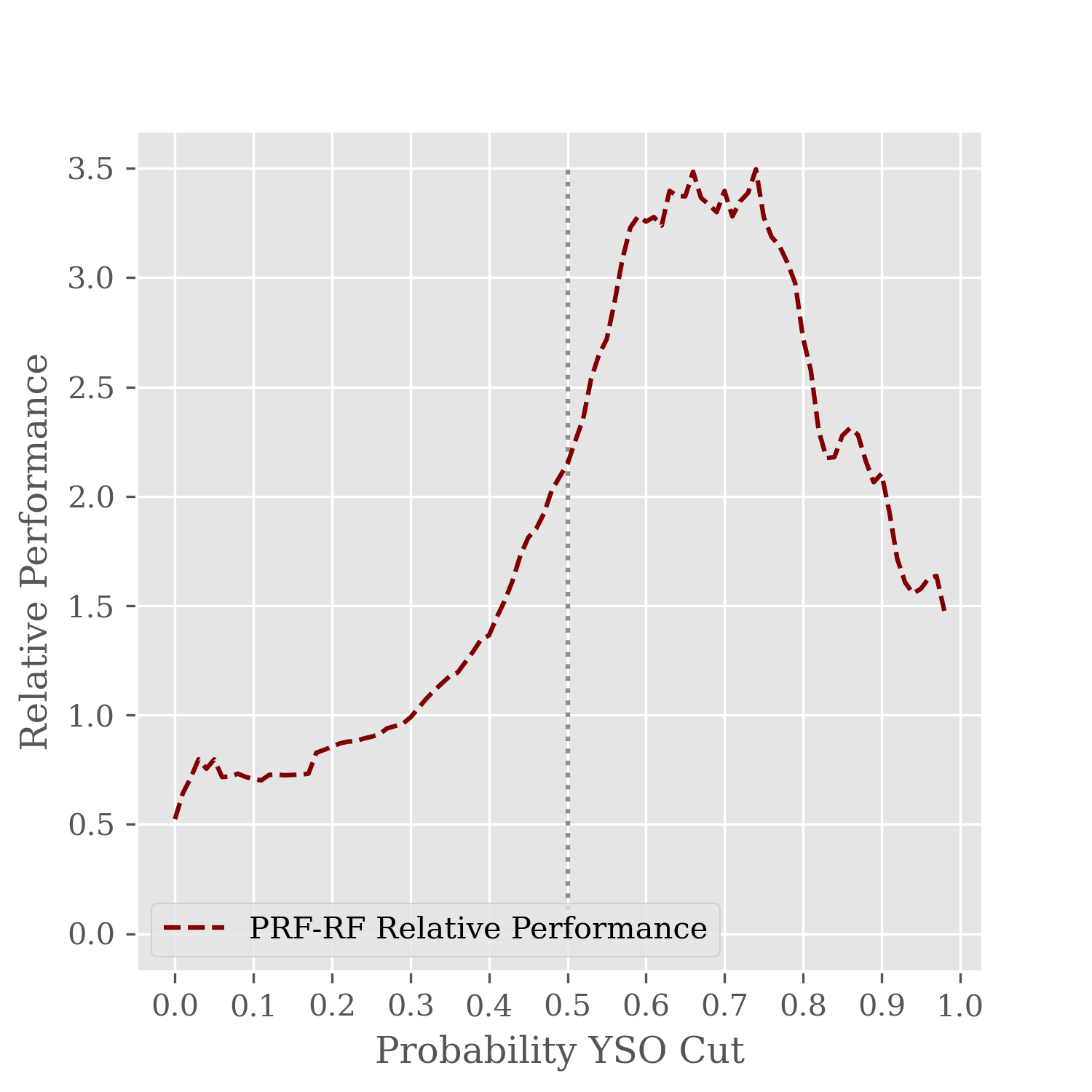}
    \caption{The relative performance of the PRF vs the RF models, see text for details on this metric. The vertical gray line marks the 50\% threshold line. }
    \label{fig:rel_perf}
\end{figure}

{We find that the PRF is able to outperform the RF when looking for True Positives. Figure~\ref{fig:rel_perf} shows the relative performance based on the number of True Positives returned as a function of precision. This metric can be determined from rearrangement of the precision equation to find Equation~\ref{eq:rel_perf}, which is what we plot as ``Relative Performance'' (RP), i.e., }
~
\begin{equation}
    \mathrm{RP} =
    \frac{\mathrm{TP}_{PRF}}{\mathrm{TP}_{RF}}= \frac{\mathrm{P}_{PRF}}{\mathrm{P}_{RF}} \frac{\mathrm{N}_{PRF}}{\mathrm{N}_{RF}.}
    \label{eq:rel_perf}
\end{equation}
Hence, the higher number of cYSOs found with the PRF is due to a higher number of True Positives. Indeed, at a threshold of 50\%, RP indicates that we find  {$> 2.0$ times} as many True Positives with the PRF as the RF.

\begin{figure}
    \centering
    \includegraphics[width=\columnwidth]{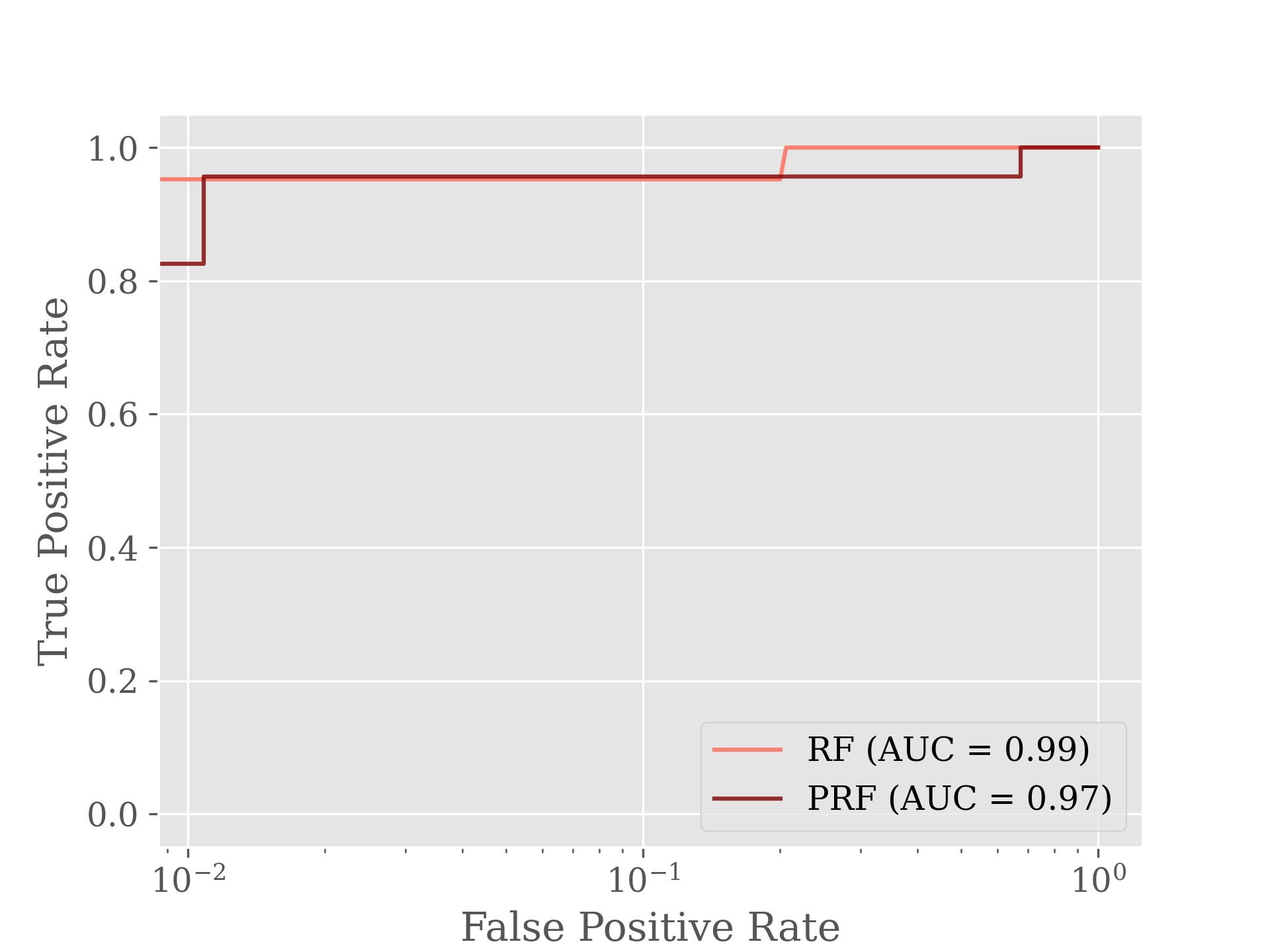}
    \caption{{ROC curves for each model with threshold based on the probability of an object being a YSO.}}
    \label{fig:roc_curve}
\end{figure}

We further address standard machine learning metrics in addition to those previously discussed.
Figure~\ref{fig:roc_curve} shows the Receiver Operating Characteristic (ROC) curve (with logarithmic abscissa as is standard for imbalanced classification) for the identification of cYSOs with a threshold determined by the probability of an object being a YSO. 
The ROC curve is a standard metric for general machine learning applications.  The Area Under the Curve (AUC) here provides the goodness of fit, with an AUC value approaching 1 being an excellent fit. We include this Figure, however, to demonstrate the deceptive nature of the AUC when applied to imbalanced data sets.  For example, if almost all objects are placed in a large class, the AUC value will approach 1 regardless of the performance of the much smaller class.

 {To explore further the robustness of the YSO classifications to the input data,} we tested several different  {filter combinations.  Beyond using all six available NIRCam bands, we also performed six other tests, where in each test just five NIRCam bands were used, in order to explore the relative impact of a dropped band on the classifications. }
 {The scores remained similar with these tests, though the exact objects in the full dataset that are identified as YSOs varied from as few as 496 (when the F187N filter data were removed) to as many as 1669 (when the F444W filter data were removed). Based on these results,} we chose to use all available bands for the models,  {but imposed a criterion that any source identified as a YSO at a probability $>50$\% also must be identified as such in at least four out of the six other tests where data from one band were removed. With this extra criterion applied, we retrieve a total of \totysosfrac{} objects identified as YSOs. These objects are henceforth referred to as ``candidate YSOs" (cYSOs).}

 {Due to the nature of how objects are classified in a RF algorithm, we emphasize that a 50\% probability of an object being a YSO does not guarantee that said object is a YSO, but rather implies that half the objects classified to such degree are indeed contaminants. By running the PRF multiple times with varying input data (700 runs in total of 100 decision trees each), however, we are varying our decisions by sufficient degree to minimize this effect. Indeed, if we estimate our False Positive rate from those objects with probability $>50\%$ (no extra criterion added), we find $\sim226$ objects are statistically likely to be False Positives. Subtracting this from our total count of cYSOs implies a final count of 459 True Positive YSOs, while, when we impose the additional criterion, our final number of cYSOs is \totysosfrac{}, and thus we conclude we are reasonably minimizing our False Positives. }
 {Furthermore, }we estimated earlier that the number of YSOs likely to be within the Cosmic Cliffs region is approximately 630, using estimates from \citet{Preibisch2014xray}, and we find our new count of cYSOs within the region is well within this estimate.

\begin{figure}
    \centering
    \includegraphics[width=0.8\columnwidth]{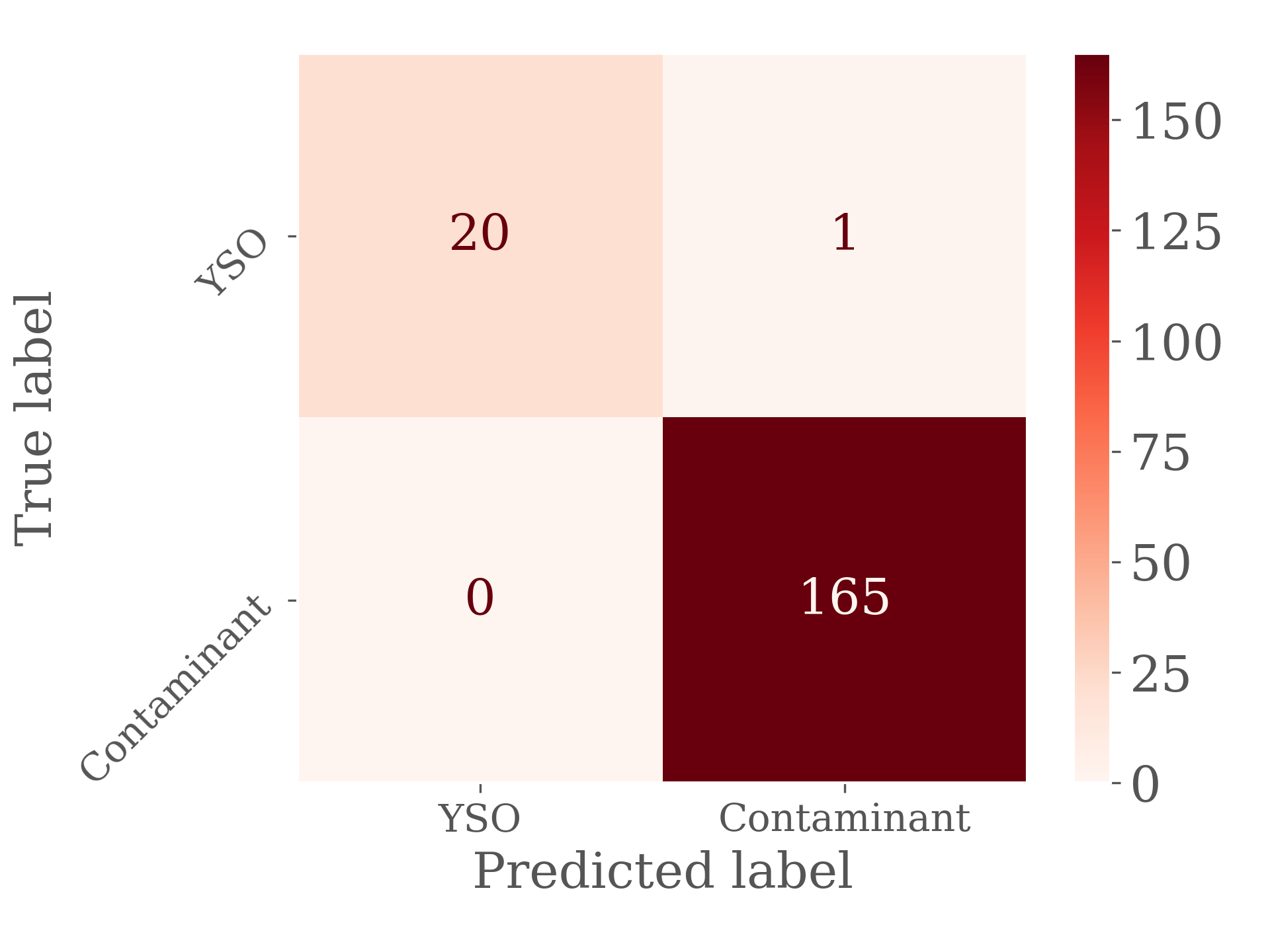}
    \includegraphics[width=0.8\columnwidth]{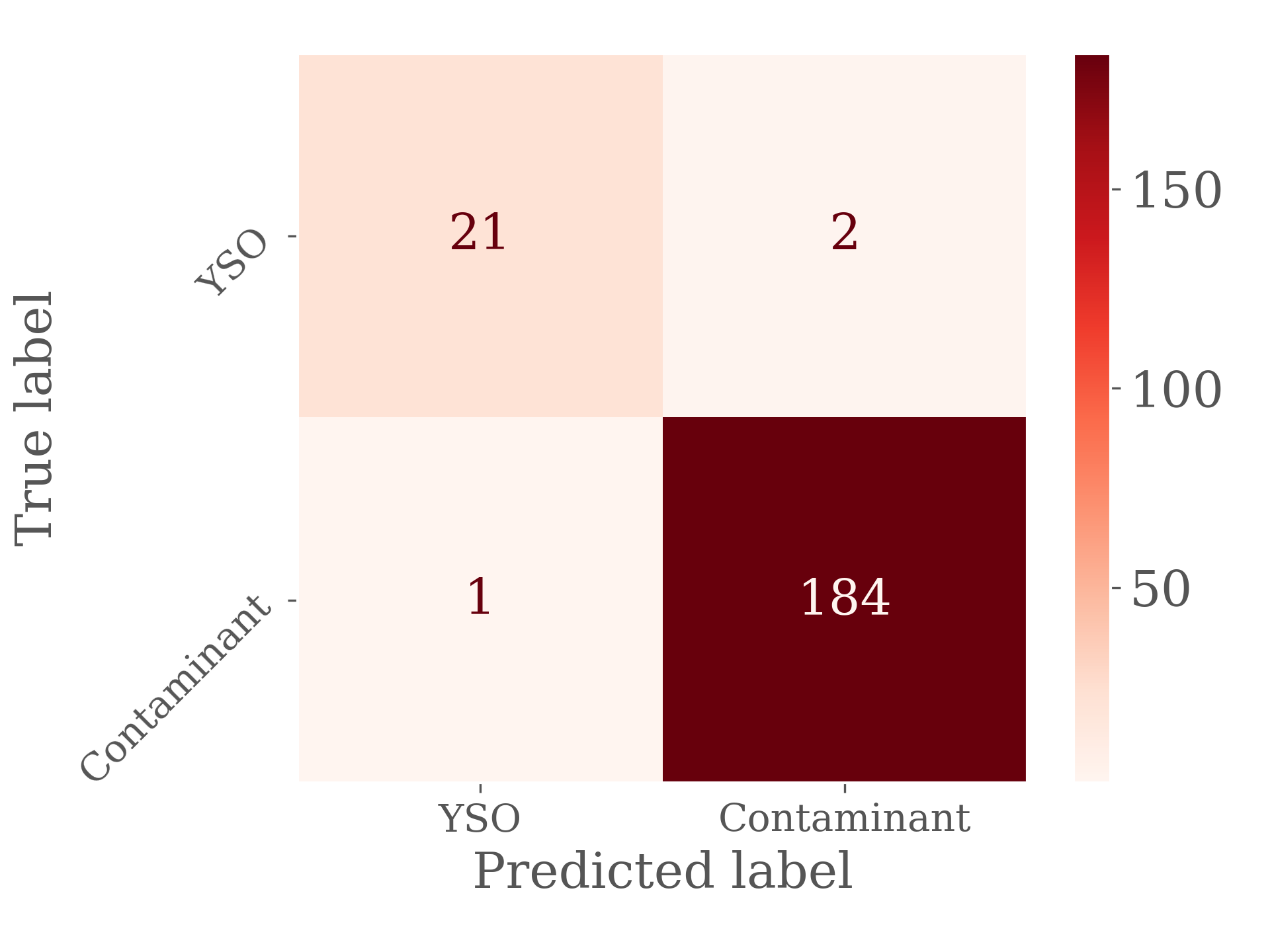}
    \caption{{The confusion matrices for the validation set using a cut of 50\% probability for an object to be a YSO based on the RF results (top) and PRF result (bottom).}}
    \label{fig:cm}
\end{figure}

 {To evaluate the goodness of the results once the extra criterion has been applied,  Figure~\ref{fig:cm}
shows the confusion matrices for the validation set using both RF (top) and PRF (bottom) models. 
The metrics are indeed identical when only considering objects found in all bands, i.e., the PRF and RF perform at the same level when tested on the same data. The original dataset used \ysotrain{} YSOs and \conttrain{} contaminants to build out an augmented dataset. Therefore, both algorithms are classifying the same new data (\ysotest{} YSO and \conttest{} contaminants) to the same extent, while the PRF is additionally classifying \totadtest{} objects with missing data (\ysoadtest{} YSOs and \contadtest{} contaminants).}
{Both the RF and PRF methods produce high scores, and so we instead refer to Figure~\ref{fig:rel_perf} to illustrate the relative performance of these algorithms.}

\begin{figure}
    \centering
    \includegraphics[width=\columnwidth]{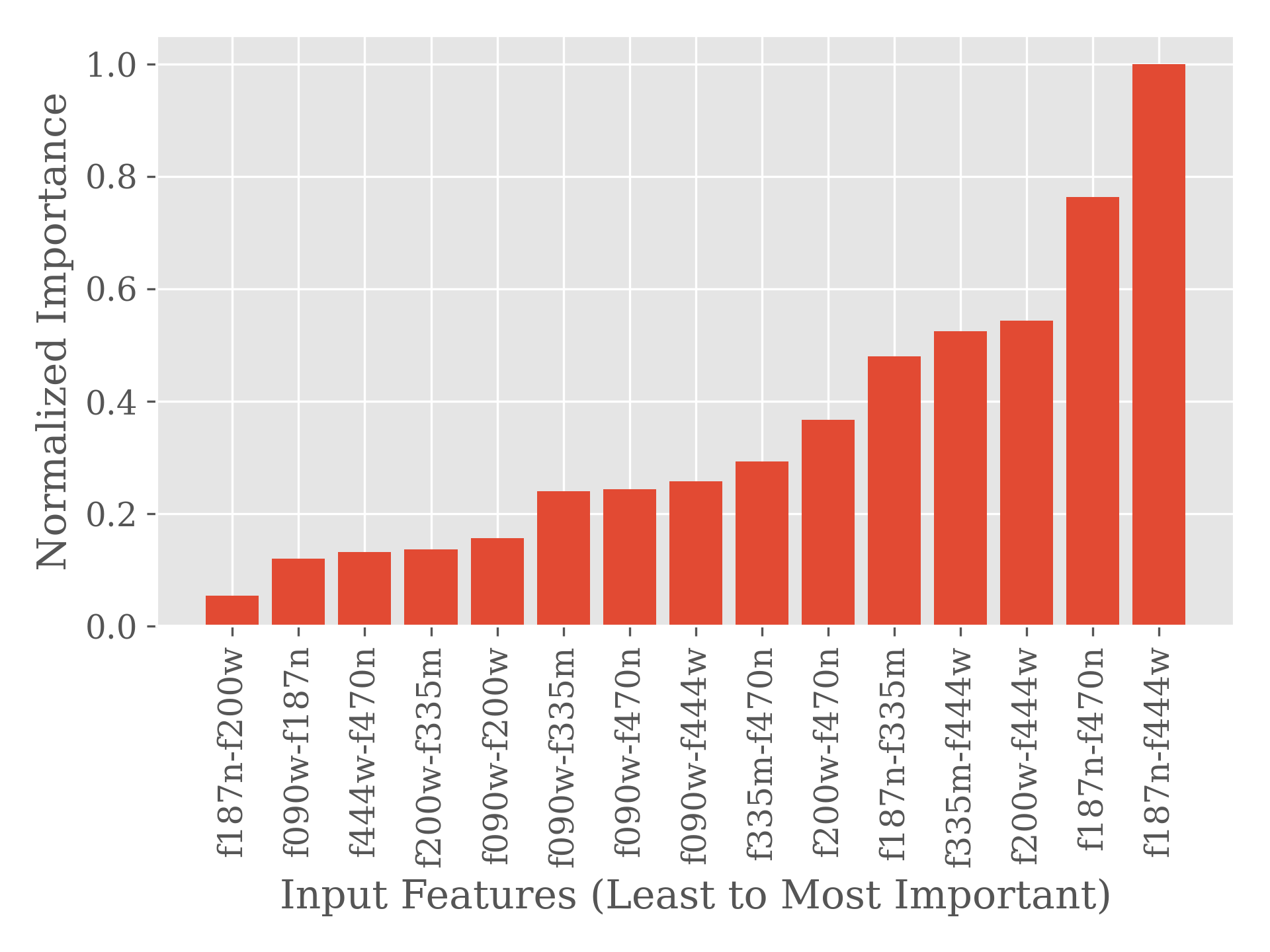}
    \caption{The relative importance of each color to the classification model, based on the \texttt{feature\_importances\_} metric of the PRF that uses all available data.}
    \label{fig:feat_import}
\end{figure}

 {Finally, we determine the usefulness of the different filters in this work. 
Figure~\ref{fig:feat_import} shows the feature importance calculated over 700 runs, 100 with all bands included and 100 each with one band removed. Feature importance is calculated during runtime by the PRF algorithm, and indicates which features are weighted most highly when making classification decisions. We find that, of our fifteen total features, colors involving the F444W  and F187N band are the most important, with those two bands appearing three times each in the top five features, followed by F335M, F200W, and F470N. F090W appears to be the least important for identifying YSOs, as might be expected given its relatively short wavelength.}

\section{Results} \label{sec:Results}

From our photometry of the JWST Cosmic Cliffs field, we retrieve a total of \totobjs{} objects. Of these, \totysosfrac{} are cYSOs, i.e., they have a probability greater than 50\% of being so based on the PRF model with all bands included, { as well as in 4 out of 6 tests where in each test the data from one filter are excluded.}
This total includes \totnewysoscompspitz{} cYSOs not previously identified from Spitzer data, of which \totprfysos{} were identified only by the PRF as they contained missing data and hence could not be classified by the RF model. \totprfrfysosfrac{} of the remaining cYSOs are found to be cYSOs with the regular RF  {with the same criterion,} and \totprfnotrfysosfrac{} are not found to be cYSOs with the RF to this threshold. In our catalog, we mark the latter \totprfnotrfysosfrac{} as being insecurely classified. 

{These numbers are an underestimate as we are not thoroughly sampling the more evolved Class III YSOs, or the more embedded Class I and Class 0 YSOs, both due to the nature of using only NIR data and the effects of our limited training set.  As in all observational studies, we are of course also limited by the finite resolution and sensitivity of the instrumentation used, in this case JWST.}
Table~\ref{tab:big_tab} lists the J2000 coordinates and JWST magnitudes for a sample of the cYSOs found with our final model.

\begin{deluxetable*}{cccccccccc} 
\tablecaption{A sample of the fluxes, J2000 co-ordinates, and probabilities for sources classified as cYSOs. \label{tab:big_tab}} 
\tablehead{\colhead{R.A.}&\colhead{decl.}&\colhead{F090W}&\colhead{F187N}&\colhead{F200W}&\colhead{F335M}&\colhead{F444W}&\colhead{F470N}&\colhead{Prob PRF}&\colhead{Frac. Runs}} 
\startdata 
159.2181&-58.6358&24.56&15.32&14.26&10.87&8.40&8.84&1.00&1.00\\ 
159.2588&-58.6160&19.29&16.23&16.15&14.50&13.51&13.38&1.00&1.00\\ 
159.2987&-58.5735&19.32&15.85&15.59&14.14&13.38&13.29&1.00&1.00\\ 
159.2486&-58.6128&23.74&20.48&20.20&18.74&17.90&17.84&1.00&1.00\\ 
159.2503&-58.5912&16.60&14.22&13.20&12.20&11.45&11.48&1.00&1.00\\ 
159.2360&-58.6164&18.36&15.04&14.68&13.49&12.71&12.69&1.00&1.00\\ 
159.2413&-58.5922&20.78&17.75&17.37&16.01&15.15&15.01&1.00&1.00\\ 
159.2462&-58.5902&16.84&14.85&14.07&13.16&12.35&12.29&1.00&1.00\\ 
159.2530&-58.5958&18.18&15.33&14.39&13.47&12.78&12.80&1.00&1.00\\ 
159.2106&-58.6312&19.24&15.34&14.76&13.31&12.25&12.20&1.00&1.00\\ 
159.2780&-58.5721&15.93&13.22&12.90&11.56&10.77&11.17&1.00&1.00\\ 
159.2434&-58.6053&19.07&15.71&15.30&13.63&12.51&12.42&1.00&1.00\\ 
159.2582&-58.6283&19.03&16.09&15.77&14.37&13.50&13.46&1.00&1.00\\ 
159.2768&-58.5883&15.55&13.58&12.74&11.68&10.87&11.46&1.00&1.00\\ 
159.2036&-58.6346&15.22&12.66&10.71&11.10&10.14&10.66&1.00&1.00\\ 
159.2188&-58.6237&16.88&14.61&14.00&13.38&12.62&12.61&1.00&1.00\\ 
159.2871&-58.5967&23.12&20.40&19.94&18.76&17.89&17.62&1.00&1.00\\ 
159.2750&-58.5734&20.95&17.94&17.59&16.03&15.07&14.94&1.00&1.00\\ 
159.2221&-58.6317&17.14&13.20&11.87&11.57&10.35&10.54&1.00&1.00\\ 
159.2453&-58.6284&18.89&15.75&15.38&13.92&13.08&13.10&1.00&1.00\\ 
159.2150&-58.6062&23.21&19.66&19.49&18.21&17.28&17.13&1.00&1.00\\ 
159.2850&-58.6152&16.23&13.84&13.63&12.66&12.05&12.19&1.00&1.00\\ 
159.2536&-58.6063&16.88&14.38&14.16&13.31&12.64&12.62&1.00&1.00\\ 
159.2514&-58.6416&16.60&14.37&14.03&13.17&12.57&12.63&1.00&1.00\\ 
159.2789&-58.6154&14.86&13.22&12.73&12.34&11.93&12.23&1.00&1.00\\ 
159.2399&-58.6435&22.21&19.97&19.47&18.76&17.93&17.92&1.00&1.00\\ 
159.2051&-58.6393&20.80&14.69&13.25&12.34&11.85&12.19&0.99&1.00\\ 
159.2972&-58.5793&15.85&13.12&13.09&11.93&11.10&11.42&0.99&1.00\\ 
159.2448&-58.5884&18.01&15.80&15.00&13.69&12.98&12.97&0.99&1.00\\ 
159.2181&-58.6067&19.84&17.42&17.31&16.25&15.61&15.72&0.99&1.00\\ 
159.2250&-58.6221&22.44&14.92&14.11&11.66&9.75&10.26&0.99&1.00\\ 
159.2383&-58.6359&24.02&22.08&20.99&20.39&19.20&19.17&0.98&1.00\\ 
159.2889&-58.5695&22.44&19.85&19.67&17.97&16.54&16.43&0.98&1.00\\ 
159.2487&-58.5951&24.07&22.02&21.18&20.33&19.61&19.71&0.98&0.83\\ 
159.2233&-58.6029&21.14&17.72&17.38&15.81&15.14&15.07&0.98&1.00\\ 
159.2668&-58.5768&22.85&19.98&19.47&18.10&17.51&17.19&0.98&1.00\\ 
159.2593&-58.6077&21.11&18.31&18.19&16.84&15.94&15.84&0.97&1.00\\ 
159.2429&-58.6380&21.93&20.24&19.12&18.16&17.47&17.46&0.97&0.83\\ 
159.2789&-58.5893&21.43&19.56&18.91&17.74&17.08&16.83&0.97&1.00\\ 
159.2693&-58.6017&22.50&20.22&19.52&18.59&17.90&17.77&0.97&0.83\\ 
\enddata 
\tablecomments{A probability less than 0.50 meant the object was not classified as a YSO in that algorithm.  {Frac. Runs specifies the fraction of runs (out of six) that the object was classified as a cYSO in. }
}
 \end{deluxetable*}

\subsection{Comparison to Previous Works}


{We summarize here the similarities and differences between our work and previous studies. In particular, we look at previously identified YSOs to determine if the PRF model was indeed  unable to generalize well enough to locate them or if they were not bona fide YSOs. The former case occurs in  {four} different instances: (1) the YSO is identified by x-ray emission and is likely a Class III YSO, which can be similar to regular field stars in the NIR; (2) the YSO is identified by Herschel data in the FIR, and is not identifiable in a significant enough number of bands in the NIR to warrant classification; (3) the YSO has too many saturated pixels, yielding poor photometric measurements of its flux {; and (4) our conservative extra criterion marks the object as being not a cYSO despite having greater than 50\% probability of being so}.  The case of a previously identified object not being a bona fide YSO, after all, is caused by one of three options: (1) the source found previously was actually a blending of several sources, resulting in inaccurate SEDs; (2) the source was a reddened galaxy or evolved star rather than a YSO; or (3) the wrong source was identified as a YSO -- this last possibility is relevant only for the suggested driving sources of outflows identified by \citet{Reiter2022}.}

\begin{deluxetable*}{cccc}
\tablecaption{Candidate YSOs from other works matched to our candidate YSOs. 
       \label{tab:yso_c_table}}
\tablehead{\colhead{cYSO}&\colhead{Previous Works}&\colhead{Prob}&\colhead{Nearby}}
\startdata
J103652.3-583809&HH 1219, SPICY 7434,  {OHL*}&1.00& - \\ 
J103702.1-583658&HH c-4, SPICY 7467&1.00& - \\ 
J103647.3-583810&MHO 1634, SPICY 7423, OHL*& {0.63}& - \\ 
J103650.5-583752&MHO 1637, OHL*&1.00& - \\ 
J103652.3-583809&MHO 1638& {0.01}&1.00 \\ 
J103653.8-583748&MHO 1639, HH 1221, HH 1003 Aa& {0.78}& - \\ 
 {J103651.3-583709}&MHO 1640& {0.52}& {0.87} \\ 
J103654.4-583618&MHO 1645, MHO 1646& {0.66}& - \\ 
J103654.0-583720&MHO 1647, HH 1002a, SPICY 7441,  {OHL*}&0.99&- \\ 
J103652.9-583737&MHO 1650&0.01& {0.70} \\ 
J103653.3-583754&MHO 1651, HH 1003B/Ca, SPICY 7438, OHL*&1.00& - \\ 
J103652.7-583805&MHO 1652&0.67& - \\ 
 {J103649.0-584010} &  {OHL*} &  {0.67} & -\\
J103649.2-583821&OHL*& {0.99}& - \\ 
J103648.9-583805&SPICY 7428,  {CXOU J103648.8-583804}, OHL*&1.00& - \\ 
J103652.5-583725&SPICY 7435, OHL*&1.00& - \\ 
J103656.6-583659&SPICY 7444& {0.66}& - \\ 
J103658.4-583619&SPICY 7448&1.00& - \\ 
J103659.1-583525&SPICY 7454&1.00& - \\ 
J103700.1-583528&SPICY 7461&1.00& - \\ 
J103700.3-583830&SPICY 7462,   {CXOU J103700.4-583829}&1.00& - \\ 
J103700.7-583545&SPICY 7464&1.00& - \\ 
J103700.9-583623&SPICY 7465&1.00& - \\ 
J103706.4-583518&SPICY 7475, OHL*&1.00& - \\ 
J103706.7-583420&SPICY 7476&1.00& - \\ 
J103708.4-583655&SPICY 7479&1.00& - \\ 
J103711.3-583445&SPICY 7481,  {CXOU J103711.4-583442}, OHL*&1.00& - \\ 
J103711.7-583424&SPICY 7482&1.00& - \\ 
J103652.4-583627& {CXOU J103652.5-583627}&-& {0.81} \\ 
J103653.9-584055& {CXOU J103653.9-584054}& {0.64}& - \\ 
 {J103654.8-583733}&CXOU J103654.7-583731&-& {0.64} \\ 
 {J103656.3-583705} &  {CXOU J103656.2-583706} & - &  {0.77} \\ 
J103658.7-583518& {CXOU J103658.8-583517}&-& {0.99} \\ 
 {J103700.6-583807} &  {CXOU J103700.6-583807} &  {0.78} & -\\ 
J103704.1-583500& {CXOU J103704.0-583501}&-& {0.81} \\ 
J103705.9-583518& {CXOU J103705.9-583520}&-& {0.78} \\ 
    \enddata
\tablecomments{The different catalog \citep{Ohlendorf2013,Preibisch2014xray,Kuhn2021} matches and outflow \citep{Reiter2022} numbers associated with each object are labelled, along with the probability of the object originally selected as the cYSO (within 0.0001 degrees of the position specified in previous works or matched by eye). If this probability is less than 50\%  {or not found in at least 4 of 6 runs}, we also check for nearby (within FWHM of original detector) sources of higher probability due to resolution differences in previous surveys. The J-2000 designation is marked as the object with greater probability in these cases. }
\end{deluxetable*}

{In total, 60 YSOs in the Cosmic Cliffs identified in previous works 
\citep{Ohlendorf2013,Preibisch2014xray,Kuhn2021,Reiter2022}
were matched to JWST objects within the FWHMs of their respective observations. These include 26 x-ray emitting sources, of which we expect some overlap, see Section~$\S~\ref{sec:xray}$ for more detail.
Of the 60 YSOs from previous works, 28 were matched to our cYSOs, leaving 32 objects that require a closer look to understand why the algorithm did not categorize them as cYSOs.}
{Table~\ref{tab:yso_c_table} 
lists all the cYSOs found in this work that match to YSOs identified previously, along with their PRF-based probabilities.}

\begin{figure*}
    \centering
    \includegraphics[width=\textwidth]{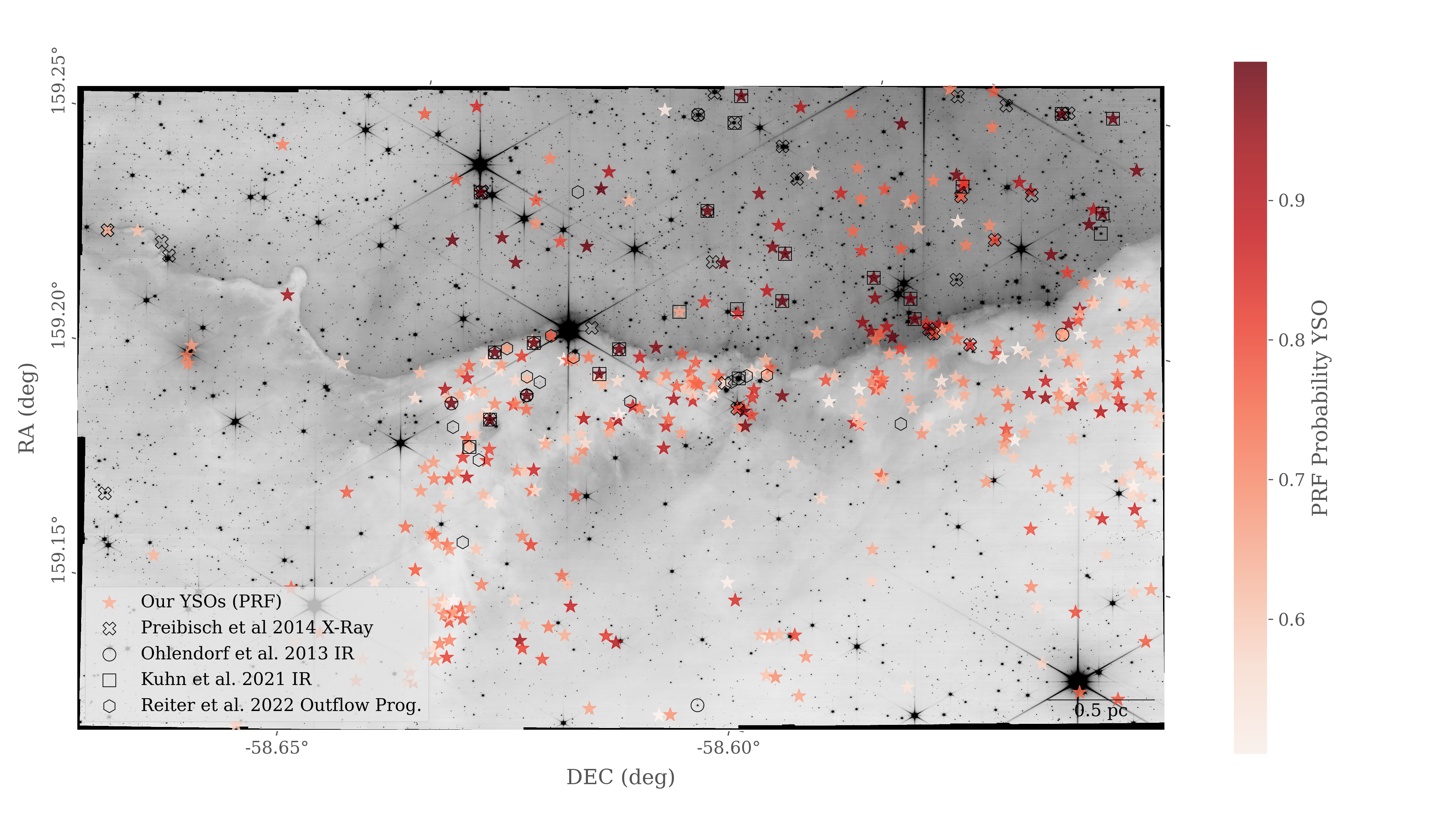}
    \caption{{All cYSOs in this region matched to Webb sources from \citet{Preibisch2014xray} (black X's), \citet{Ohlendorf2013} (black circles), \citet{Kuhn2021} (black squares),  \citet{Reiter2022} (black hexagons), and 
    with our own cYSOs marked in varying shads of red based on their probability of being a YSO.}}
    \label{fig:colimage}
\end{figure*}

Given differences in the angular resolution and sensitivity between the JWST data and previous works, we chose as the most likely corresponding source to be the one within the FWHM of the lower resolution instrument. If multiple sources are located within this FWHM, however, the brightest source was nominally selected.
After classifying the data, however, we found that 6 of the 60 matched sources did not have high probability of being YSOs, though other sources within the original FWHM did. In such cases, we marked the object with greater probability as a `nearby' possible match in Table~\ref{tab:yso_c_table}, and updated the J2000 designation of the match to that of the most YSO-like object within the FWHM that we then categorize as a cYSO. After this change, we now have matches to  {34} out of 60 cYSOs. Evaluation of other possible driving sources of outflows from \citet{Reiter2022}, $\S$\ref{sec:outflows}, has this number increase to  {37} out of 60 YSOs. This rate of recovery may imply that $\sim$ {280} additional cYSOs within the field remain unidentified as cYSOs. Accounting for this number would result in a total cYSO number of  {730},  {over} the estimate of 630 as computed in \S\ref{sec:target_area} {, though this estimate does not include sub-stellar objects, see following Discussion. }Figure~\ref{fig:colimage} shows all cYSOs identified with the PRF, as well as the YSOs found in other works that are matched to JWST sources \citep{Ohlendorf2013,Preibisch2014xray,Kuhn2021,Reiter2022}.

\subsubsection{X-ray sources}\label{sec:xray}

{\citet{Preibisch2014xray} found  {51} x-ray objects within the Cosmic Cliffs, of which only 26 were matched to JWST objects within the FWHMs of the x-ray sources, including  {five} cYSOs. A further  {six} objects have cYSOs as their secondary match.  
This small quantity of matches is unsurprising, as x-rays most effectively sample Class III YSOs \citep{Feigelson2018}, which cannot be well identified by NIR colors \citep[e.g.,][]{Dunham2015} {and are generally beyond the limits of our training set}. These x-ray observations also had a detection limit of $\sim0.8~M_\odot$ \citep{Preibisch2014xray}, likely leaving undetected much of the YSO population probed by JWST.
Furthermore, x-ray emission can also trace other astronomical objects such as fore- and background Galactic field stars, and extragalactic sources \citep{Broos2011,Preibisch2014xray,Feigelson2018}.  Thus, we label  {11} out of 26 x-ray sources as  {cYSOs}.} Having a low recovery of x-ray objects therefore arguably provides further support that any contamination by such non-YSOs into our selection of cYSOs is low, but also that we are likely under-sampling the more-evolved YSO population with less infrared excess, as expected. 

 {We note here that the one object classified as a cYSO by our PRF model that was noted as a contaminant in the original classification, see Figure~\ref{fig:cm}, is associated with an x-ray source (CXOU J103700.6-583807), and has 78\% probability of being a YSO.}

\subsubsection{IR sources}
{Of the 22 Spitzer sources found to be YSOs in \citet{Ohlendorf2013}, 14 are matched to a JWST object. Most of the YSOs are spread above the HII region layer or near the HII region edge, though five are within the HII region. Of these 14, two are identified as apparent clumps of nebulosity, and we labelled them as contaminants to help our model differentiate clumped gas from YSOs. Eleven out of twelve of the remaining sources are found by the PRF to be cYSOs with at least 50\% probability. The one source not found to be a YSO was badly saturated.}

All six YSOs in the Cosmic Cliffs identified by \citet{Gaczkowski2013} from Herschel data are located in the depths of the HII region, and no point sources are visible in the NIR for matching. Of the nearby point sources available, none are found to match cYSOs with our RF or PRF models. The lack of matches to Herschel data is not unexpected since Herschel sources may simply not have near-infrared counterparts, and our training set limits us to much less embedded YSOs than those found with Herschel. As such, this lack of matches with Herschel-identified YSOs moreover illustrates that we have successfully avoided labelling such dense knots of gas as cYSOs, though the inclusion of Herschel data to the ensemble in future remains a possibility. 

For their SPICY catalog, \citet{Kuhn2022} estimated that the contamination rate of YSOs, {as defined as the number of False Positives divided by the number of True Positives, is 10\%. We ended up removing eight out of 26 objects identified in SPICY from our training set due to blending, or $\sim 30\%$. Of the remaining sources, all were retrieved as cYSOs with our model. }

{Thus, we identify 22 out of 23 NIR YSOs, and none of the deeply embedded YSOs identified from FIR data. This matches the caveats of both our training set and the impact of using only NIR data for classification.}

\subsubsection{Outflow Driving Source Candidates}\label{sec:outflows}

{Of the 21 outflow driving source candidates identified in \citet{Reiter2022}, nine are immediately found to be cYSOs in our PRF-based classifications.  Of these, four were marked in the training set, and an additional five were identified by the PRF. Of these five, two were identified by SPICY but were blended sources and thus were not present in the training set.
Of the remaining twelve candidate driving sources, four do not have satisfactory photometry due to either saturation or sensitivity limits and eight are not identified as candidate YSOs in our models to any reasonable threshold.}

{The degree of cYSO matches to outflow driving source candidates found by \citet{Reiter2022} may be due to a handful of factors.
Indeed, tracing outflow driving sources may require a more nuanced approach.  For example, all Molecular Hydrogen Objects (MHOs) without associated Herbig-Haro (HH) objects identified by \citet{Reiter2022} do not have proper motions, and so the selected driving source candidate is identified as an object with IR-excess lying along the outflow axes. We hence check for cYSOs identified along these outflow axes, and find alternatives for three of the eight cases. The five sources that we do not identify as cYSOs may be false negatives, i.e., objects that should have been marked as a YSO but were not, possibly due {to the limits of our training set}. We describe the possible driving sources for each case in Appendix~\ref{app:reit}.}  

Of the 17 outflow driving source candidates for which we had photometry, 12 are identified as cYSOs. The five not identified as cYSOs are extremely faint objects.

\section{Discussion}\label{sec:Disc}


This work aims to provide classifications of cYSOs within the ERO JWST data of NGC 3324. There are \totobjs{} objects within the $\sim 7.4' \times 4.4'$ NIRCam field. Of these, \finaltotysos{} were identified as cYSOs, and the remainder as contaminants. 
In total, we identified \totnewysos{} new cYSOs, and confirmed \totconfirmedysos{} previously identified YSOs.

\subsection{Diversity of cYSOs}

\begin{figure}
    \centering
    \includegraphics[width=\columnwidth]{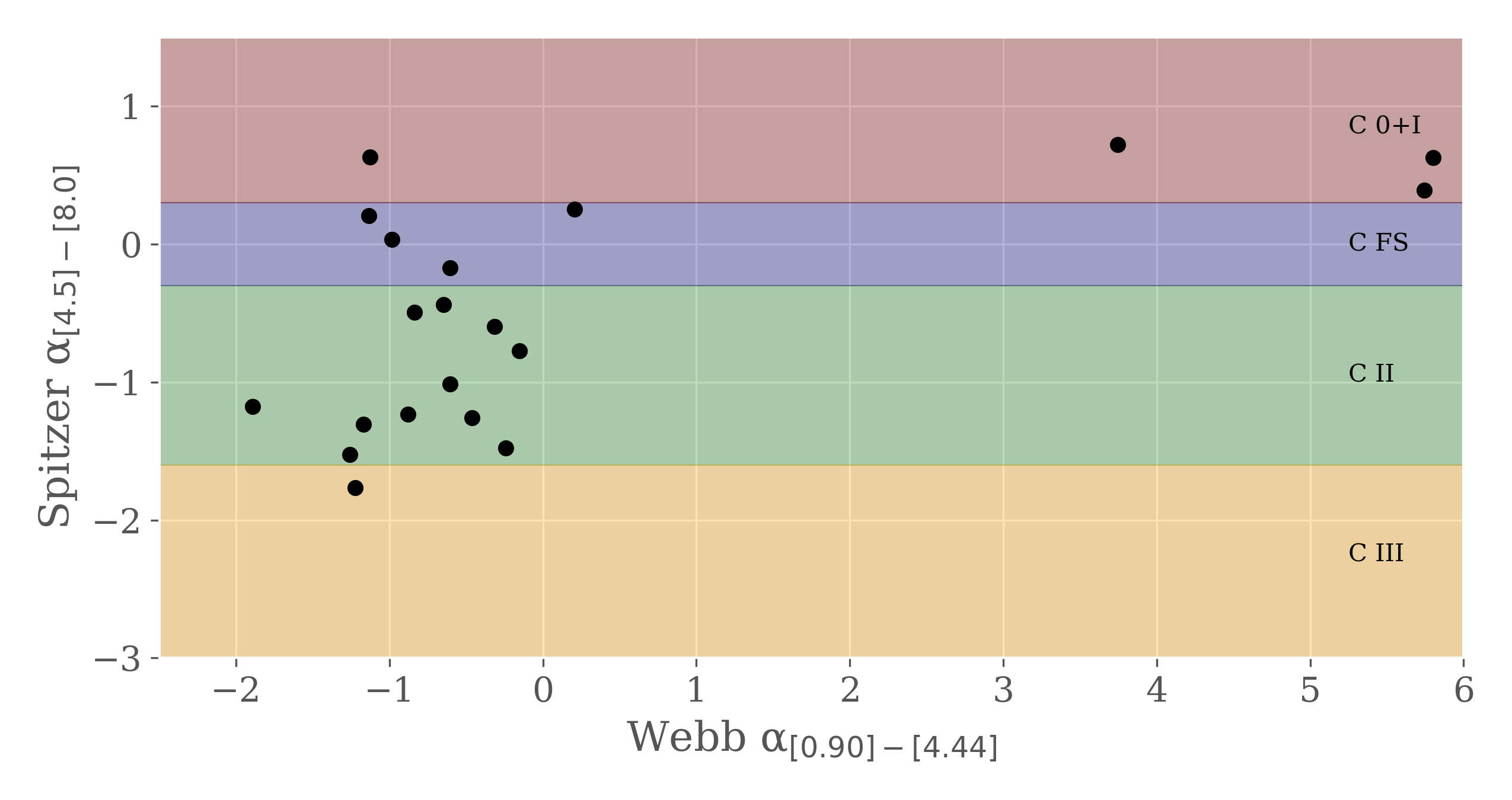}
    \caption{{Comparison of the spectral index of Spitzer identified YSOs, determined from Spitzer data (y-axis) and Webb data (x-axis). Colored segments show the spectral indices in Spitzer associated with each class, with class labels on the right.}}
    \label{fig:spect_ind_comp}
\end{figure}

The use of a machine learning method fundamentally relies upon having a training set composed of a thorough and diverse set of objects. In $\S$~\ref{sec:catcreate}, we showed that the training set covered a range of YSO stages from Class I through to Class III, based on  spectral indices calculated using Spitzer data. To approximate this index to the full set of candidate YSOs observed by NIRCam, we compute a spectral index between 0.90 $\mu m$ and 4.44 $\mu m$, i.e.,
\begin{equation}
    \alpha_{[0.90]-[4.44]} = 0.58([0.90]-[4.44])-3.57.
    \label{eq:spectral_ind_webb}
\end{equation}
This  {particular spectral index, being over a range of shorter wavelengths than available with Spitzer,} will not allow us to identify different Classes of YSOs.
Indeed, Figure~\ref{fig:spect_ind_comp} shows the lack of a strong correlation between spectral index computed from Spitzer data, and spectral index computed from Webb  {NIRCam} data.  
 {We anticipate that Class could be better determined with the inclusion of MIRI data to the ensemble. The NIRCam data, however, do allow} a comparison of the range of spectral index compared to the training set, and the trend remains of decreasing spectral index being correlated with objects of increasing Class.  {In general, $\alpha_{[0.90]-[4.44]}$ reveals the diversity of YSOs seen in the Cosmic Cliffs field.}

\begin{figure}
    \centering
    \includegraphics[width=\columnwidth]{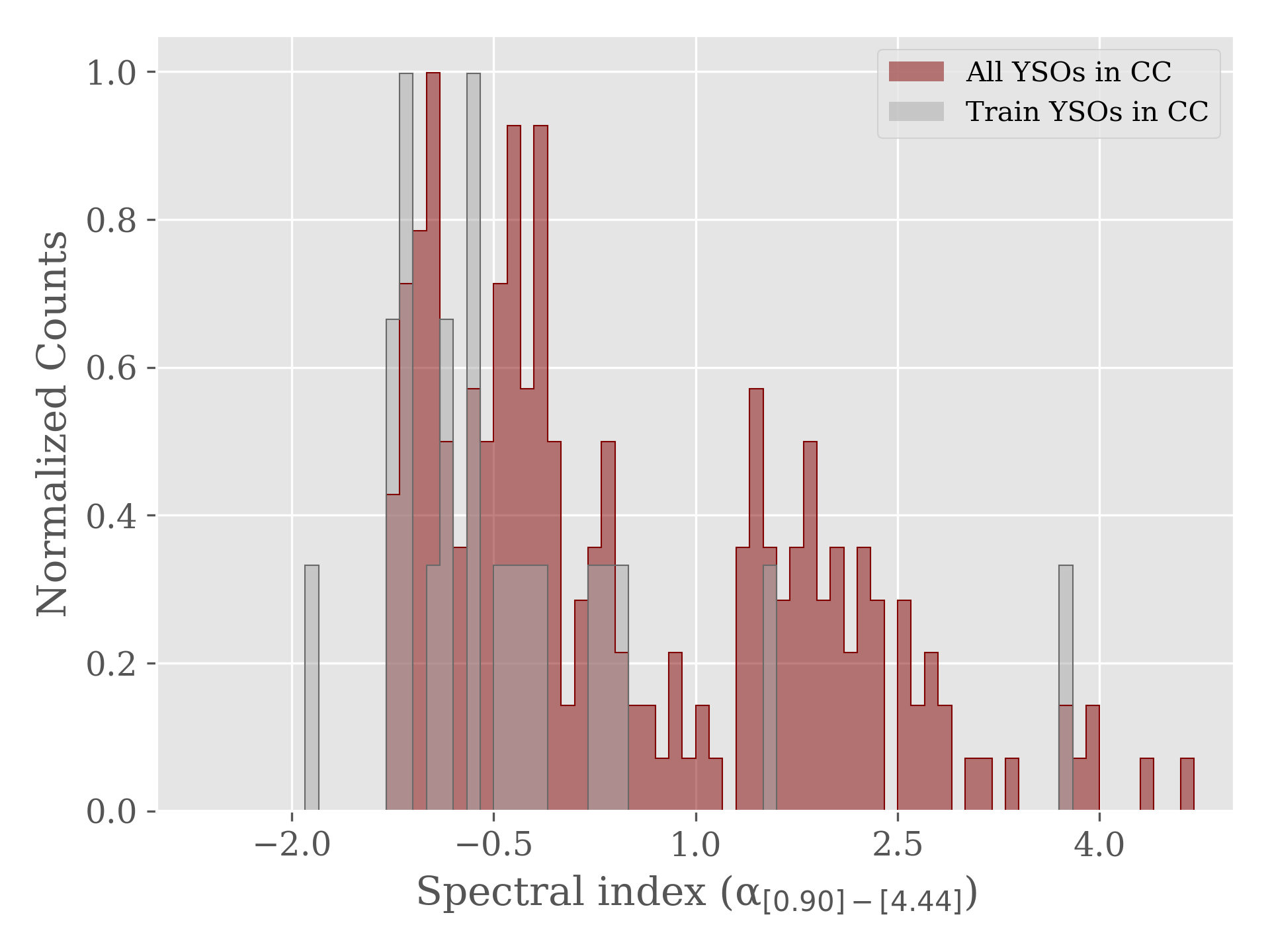}
    \caption{{Histograms of the spectral index (see Equation~\ref{eq:spectral_ind_webb}) of the full population of identified cYSOs (maroon) and the Spitzer-identified YSOs which make up the training set (gray), normalized for comparison.}}
    \label{fig:spec_ind_webb}
\end{figure}

Figure~\ref{fig:spec_ind_webb} shows a histogram of the spectral indices of all of our cYSOs, as well as a histogram of the spectral indices of the Spitzer-identified YSOs, both normalized for comparison. We see that the identified YSOs indeed cover  {most of} the full range of spectral index from those used to train the model, and in addition, that the newly identified cYSOs go beyond the Spitzer-identified YSOs towards higher spectral indices. Furthermore, this plot relies upon objects having both 0.90~$\mu$m and 4.44~$\mu$m data, while several of our identified cYSOs are missing 0.90~$\mu$m data and are located along the dustiest regions of the cloud edge. Hence, we find that we are able to identify YSOs through a broad range of spectral indices. 

{Objects of extreme spectral index, i.e., those that would be linked to clear cases of more embedded Class 0/I YSOs and Class III YSOs, are the most ill-sampled in our population. Thus, the limitations of our identified cYSOs match the caveats of using only near-infrared data to identify YSOs.}

\subsection{CMDs and CCDs} \label{sec:cmds}

\begin{figure*}
    \centering
    \includegraphics[width=\textwidth]{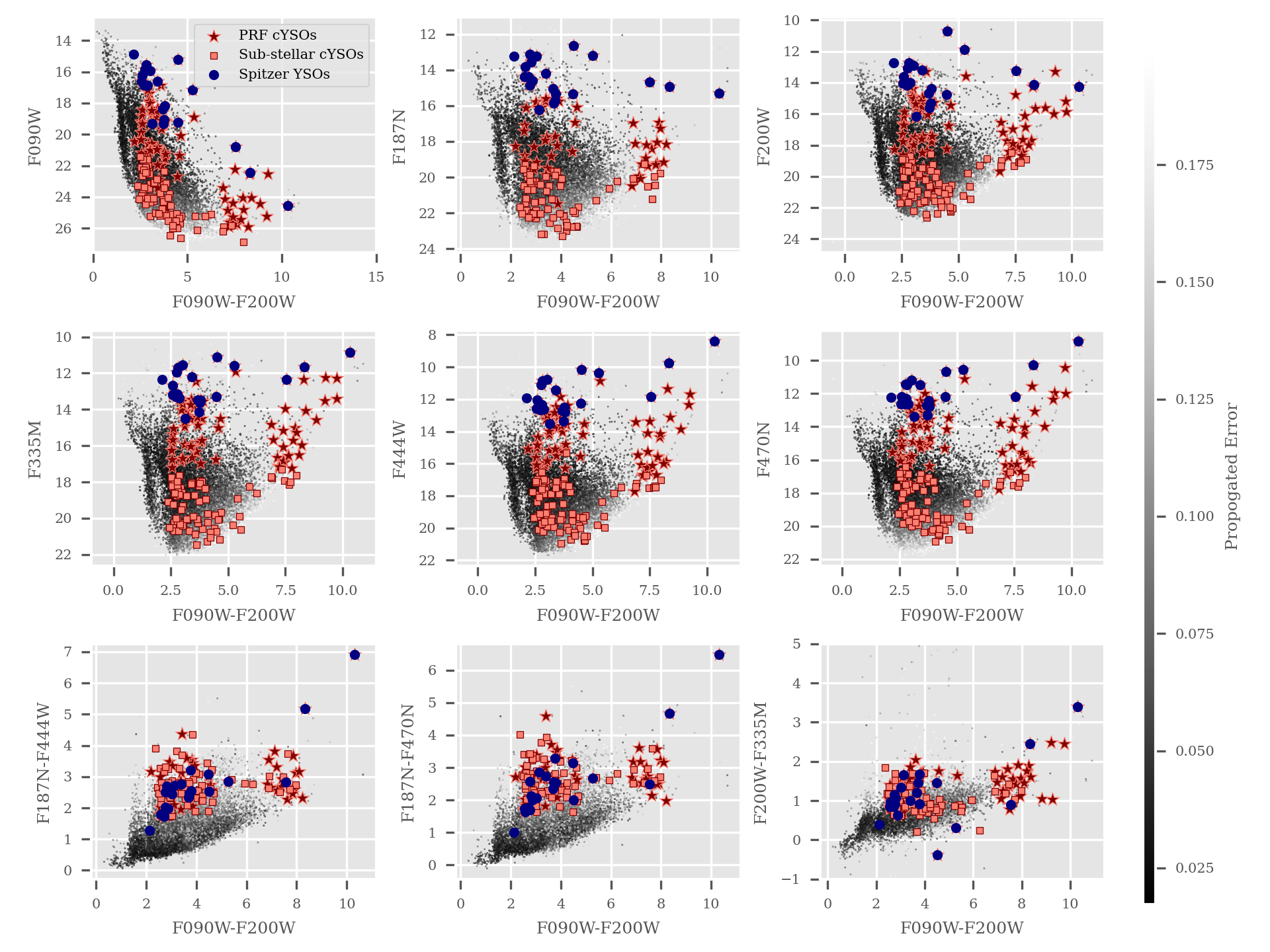}
    \caption{{Color-magnitude and color-color diagrams of cYSOs (maroon stars) {, sub-stellar cYSOs (salmon squares)}, 
    Spitzer-identified YSOs (navy circles), and contaminants (gray circles with color gradient specifying the error of the photometry). }}
    \label{fig:cmds}
\end{figure*}

\begin{figure}
    \centering
    \includegraphics[width=\columnwidth]{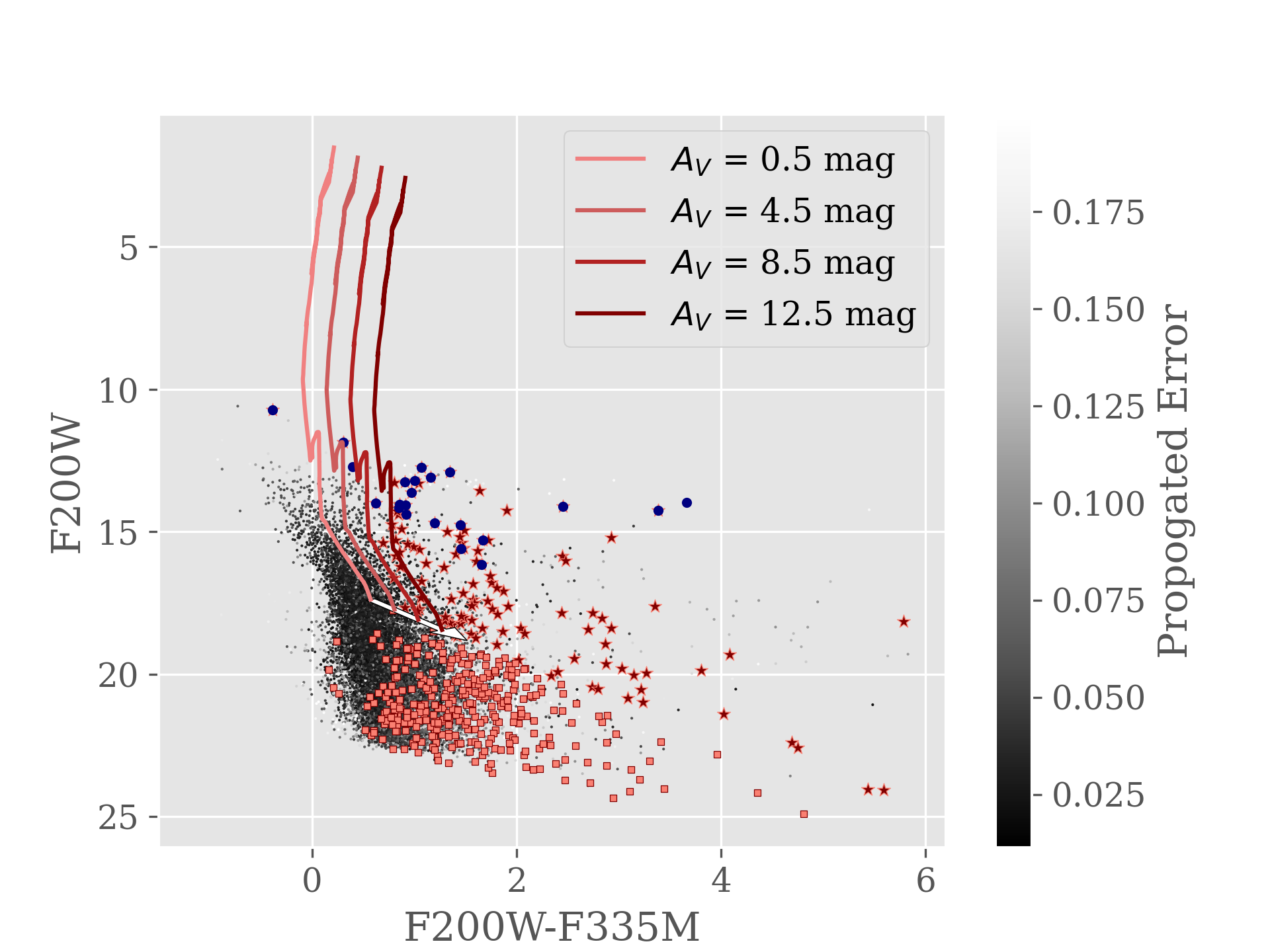}
    \caption{Diagram of [F090W]-[F200W] color vs. F200W magnitude for cYSOs and contaminants (symbols are as in Figure~\ref{fig:cmds}) overlaid with pre-main-sequence object isochrones from PARSEC models at an age of 2 Myr and various foreground extinctions.}
    \label{fig:isochrones}
\end{figure} 

YSOs have often been found based on their color-excess as determined from color-magnitude diagrams \citep[CMDs; e.g.,][]{Gutermuth2009}. As a check to see if the cYSOs we find show excess, we plot several different CMDs as well as color-color diagrams in Figure~\ref{fig:cmds}. 
The overall cYSO population remains largely on the red edges of the CMD distributions, while often following a sharp angle from the main populations in the color-color diagrams. Spitzer-detected sources are located in the brighter regions of the CMDs ($\lessapprox 14$ mag) due to the lower sensitivity of Spitzer.
The removal of magnitudes and the use of only colors in our training sample was hence necessary to avoid biasing our models to only the brightest objects.

We utilize the PARSEC evolutionary models \citep{Nguyen2022} to plot the expected pre-main sequence isochrones for YSOs in the NIRCam filters over a range of foreground extinctions and provide context to the masses of the cYSOs in the Cosmic Cliffs field.  We assume an age of 2 Myr for the pre-main sequence age, and solar metallicity, and plot in Figure~\ref{fig:isochrones} isochrones for a range of foreground extinctions from $A_V = 0.5$ to $A_V = 12.5$.  We use the extinction law from \citet{Wang&Chen2019} with $R_V = 3.1$ to determine extinction at different wavelengths.
Additionally, it is worth noting that the predicted colors and magnitudes from the PARSEC models are for the central source and do not include a disk contribution, which will further redden their colors.

The PARSEC pre-main-sequence isochrones only reach down to a mass of 0.09 $M_\odot$, and we find about \fracsubs{} of cYSOs are fainter than this cutoff over a wide range of extinctions, insinuating that we are seeing a substantial sub-stellar population in the Cosmic Cliffs field. {We note that, although this regime of the CMD is likely to contain extra-galactic sources  {in general}, 
the Cosmic Cliffs  {is itself quite near the Galactic mid-plane at -00.1633 degrees in Galactic latitude}. As such, these cYSOs are unlikely to be extra-galactic contaminants  {due to the intervening Galactic dust across the full Galactic disk}.}  {As seen below in \S\ref{sec:ysosd}, their spatial distribution across the field is not uniform, as might be expected if they were background galaxies.}  {We further looked at each of these cYSOs by eye to determine if any have the more extended morphology typical of galaxies, and all appear as point sources.} 

\cite{Scholz2022} estimate that below 15 $M_{JUP}$, brown dwarfs formed by core collapse make up only 0.25\% of the population of extremely low-mass objects, while rogue planets make up a more substantial fraction. If so, we may have also detected a substantial number of young rogue planets within our sample. 
 {We refer to this population as sub-stellar cYSOs, although we cannot conclusively determine if these are brown dwarfs or rogue planets without an associated training sample.}
Further observations are also necessary to ascertain the exact nature of the fainter cYSOs in the Cosmic Cliffs.  
 {In Figure~\ref{fig:cmds}, we distinguish the locations of the sub-stellar cYSOs in the various CMDs and CCDs.}

\begin{figure}
    \centering
    \includegraphics[width=\columnwidth]{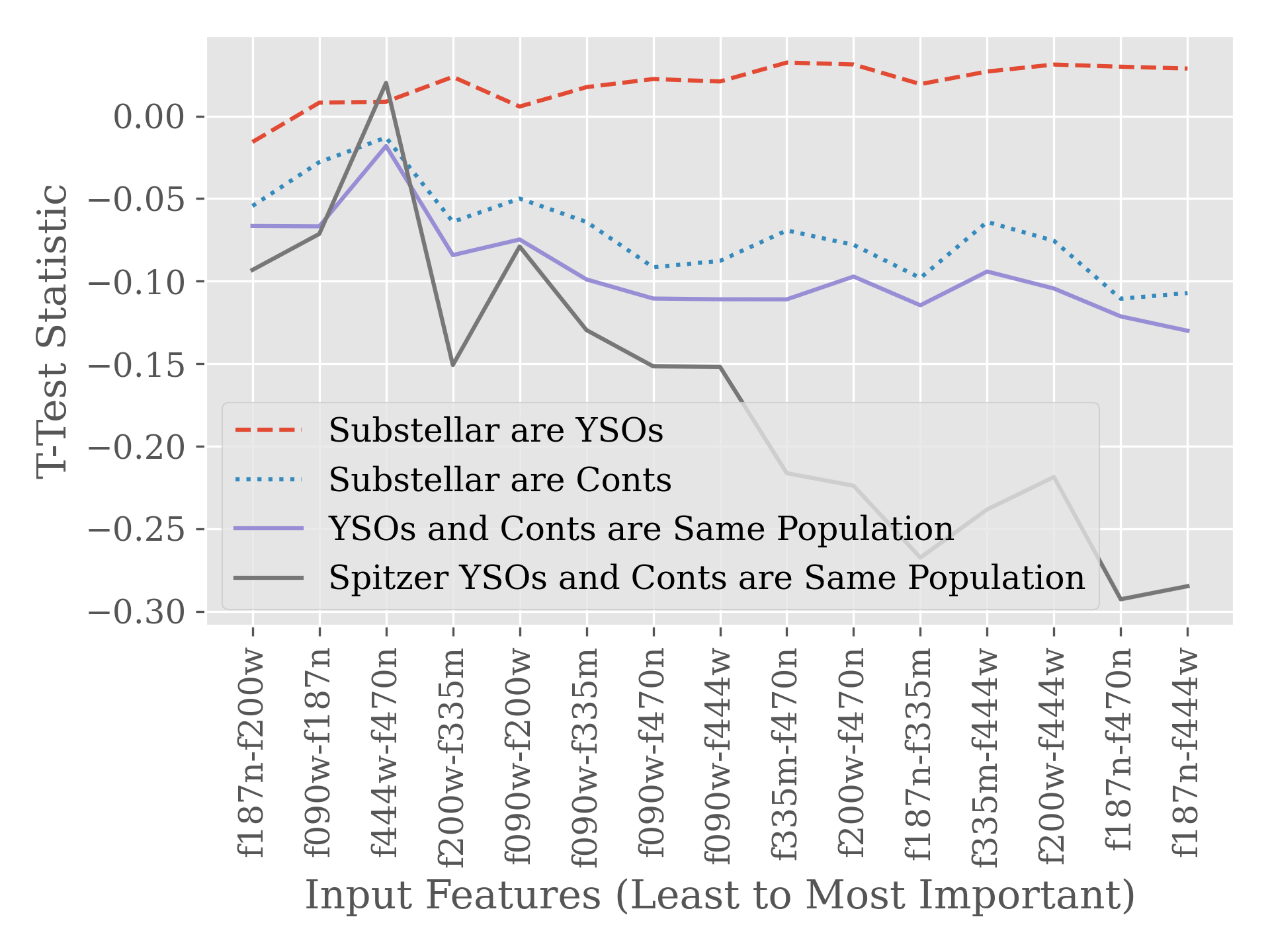}
    \caption{ {Welch's T-test investigation of the population to which the sub-stellar objects belong. In this instance ``YSOs'' are those cYSOs above the sub-stellar threshold, ``contaminants'' are either above or below the sub-stellar threshold depending on if they're being compared to the ``YSOs'' or the sub-stellar population, respectively. }}
    \label{fig:t-test}
\end{figure}

 {To test further that the young sub-stellar cYSOs indeed belong to the cYSO population, we utilize Welch’s t-test \citep{Welch1947}, as available through the Python \texttt{SciPy} library \citep{Scipy}. This test returns the t-statistic, which is a measure of the similarity of population means with variance accounted for. We compute t-statistics to determine in which population the sub-stellar cYSOs are more likely to belong: the ``regular" cYSOs or the contaminants. Figure~\ref{fig:t-test} shows these metrics, where the t-statistic is divided by the degrees of freedom to place all relationships on the same scale versus the input feature from least to most important, as determined by the PRF itself (see \S \ref{sec:PRF_apply}. }

 {From the t-statistics, we find a clear trend in similarity of population means between classes that we expect to be of the same population (e.g., regular cYSOs and the faint young sub-stellar candidates) versus the similarity between classes we expect to be of different populations (e.g., regular cYSOs and contaminants, where we only consider contaminants of similar brightness to those cYSOs). Namely, the t-statistic remains close to zero, indicative of greater similarity, for the hypothesis that the young sub-stellar candidates and regular cYSOs are drawn from the same population. This behavior is seen for all features. Conversely, the t-statistic departs drastically from zero with the hypothesis that the contaminants and young sub-stellar candidates are drawn from the same population. Therefore, the sub-stellar cYSO population is most similar, at least in population mean, to the regular cYSO population, rather than the faint contaminant population.}

\subsection{YSO Spatial Distribution}
\label{sec:ysosd}

{The spatial distribution of YSOs within Gum~31 was previously analyzed by \citet{Ohlendorf2013}. They found that this distribution was far from uniform, with many YSOs found in compact groups, including a considerable number found along the dusty cloud edge and in the heads of pillars. This behavior led them to propose two methods of ongoing triggered star formation were at work: collect-and-collapse and radiative triggering. }

In Figure~\ref{fig:colimage}, we find that the overall distribution of our cYSOs also follows the HII region boundary closely, while few cYSOs are found interior to it. Furthermore, a relative paucity of cYSOs is found to the south, i.e., on the left-hand side of the image. 
Our results support the picture that Gum 31 is indeed triggering star formation via the collect-and-collapse scenario, and that in particular we are sampling a section of the region currently undergoing it. Conversely, we only have one clear pillar in the image, with one cYSO identified within it and not on the head of it, and so we can neither support nor exclude the possibility of radiative triggering also being at work.

\begin{figure*}
    \centering
    \includegraphics[width=\textwidth]{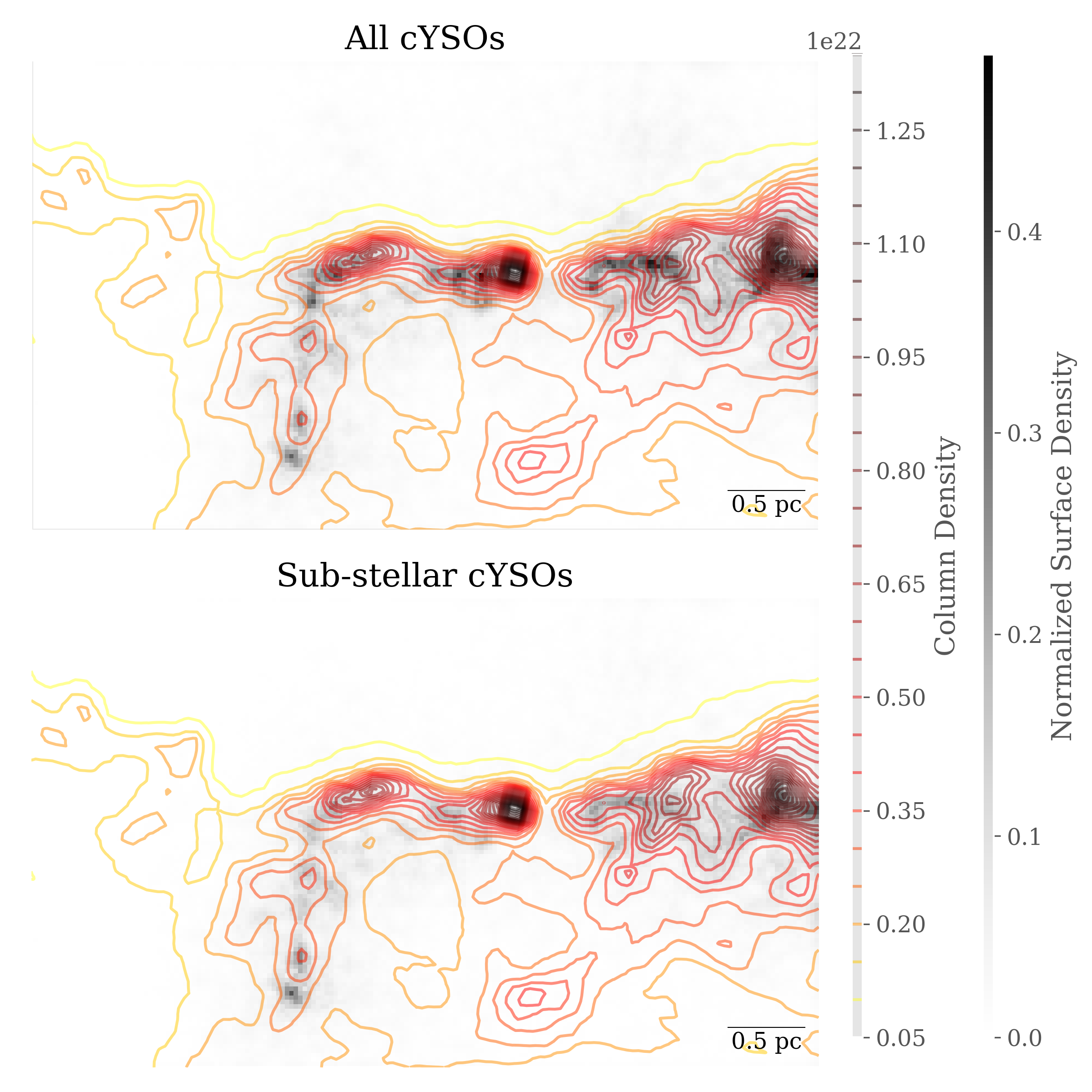}
    \caption{Binned surface density with column density contours overlaid { for the full cYSO population (top) and that of the sub-stellar cYSO population (bottom)}. 
    The surface density is normalized to the total number of objects in that bin to emphasize regions of YSO enhancement. The column density contours are laid in steps of A$_V$ = 0.5 mag. }
    \label{fig:col_surf_dens}
\end{figure*}

Figure~\ref{fig:col_surf_dens} shows the surface density of our cYSOs with respect to the H$_2$ column density of gas in this region, as obtained with Herschel continuum data using techniques described in \citet{Preibisch2012}. We place contours at levels from the minimum column density to the maximum column density that still includes a cYSO. The contour levels increase in steps of $5 \times 10^{20}$ cm$^{-2} \equiv A_V\/ = 0.5~\mathrm{mag}$ \citep[e.g., ][]{Bohlin1978}. We determine the average nearest neighbor surface density of YSOs in each grid cell, where the grid cell size is matched to the resolution of Herschel at $\sim 9.1 \times 10^{-4} ~\mathrm{pc^2}$ at this distance. We use a Monte Carlo method to incorporate the PRF probability of each of the \totobjs{} sources in the field being a YSO. The surface density is computed via 
\begin{equation}
    \Sigma_* = \frac{n-1}{\pi d_n^2} \label{eq:nnd},~~~~~~n =11,
\end{equation}
where we choose the 11th nearest object, following \citet{Pokhrel2020,Pokhrel2021}. Figure \ref{fig:col_surf_dens} 
is normalized to the total number of objects in each grid cell to highlight regions of enhanced YSO surface density.

The identified cYSOs surface densities clearly trace out the areas of high column density along the condensed HII region edge, while the Spitzer-detected YSOs \citep{Ohlendorf2013,Kuhn2021} are relegated to the less dense areas within the bubble or along the outer edges of the HII region.  
{These results are reminiscent of \citet{Gaczkowski2013}, where Herschel data identified the youngest sources along the HII region edges, while the slightly more-evolved YSOs, perhaps from a previous star formation episode, are located outside the HII region edges. The strong correlation between our YSO surface density and column density provides further support that the PRF model has successfully found younger and embedded YSOs in general. The location of these sources relative to column density also allows us to claim that JWST performs much better at detecting embedded YSOs than Spitzer could, despite only using NIRCam data (0.9-4.7 $\mu$m) in this study.}

{The collect-and-collapse scenario driven by the O-type stars in the center of Gum 31 places the bubble at an age of $\sim$2 Myr \citep{Ohlendorf2013}. Given the expansion of the Gum 31 bubble, a Class II YSO population \citep[whose lifetime is also $\sim 2$ Myr,][]{Evans2009} might be expected to have diffused widely in the observed field, with only the younger population of Class I YSOs situated more closely to their dense formation sites. We find that the cYSO population is primarily clustered tightly on both sides of the HII bubble edge. The close proximity of the cYSOs to the bubble edge suggests that either we are mostly sampling younger Class I YSOs, or perhaps that the more-evolved Class II YSOs are being pushed outwards along with the expanding bubble.}



Finally, we address the implications of the detection of a significant population of apparently sub-stellar objects in the Cosmic Cliffs.  This population brings into question earlier estimates about the yields of star-forming molecular clouds.  To demonstrate its impact, we calculate a star-formation efficiency (SFE) in two ways. (For convenience, we refer to young stellar objects both above and below the sub-stellar line as ``stars" in the SFE.)  First, we assume the average mass of a YSO is 0.5 M$_\odot$ \citep{Evans2009}, and remove the sub-stellar component from the ensemble. Second, we assume an average mass of 0.5 M$_\odot$ for those objects above the sub-stellar line, and include the sub-stellar component assuming 10\% of this mass, i.e., 0.05 M$_\odot$ for the sub-stellar component. 
We define the mass of cYSOs as $M_* = N_{PS}\times M_{PS}$, where $N_{PS}$ is the number of protostars, and $M_{PS}$ is the mass of a protostar. We then define $M_{gas}$ as $M_{gas} = (2~m_H/X)~ N(H_2)$, where $m_H$ is the mass of a hydrogen atom, and $X=0.71$ is the fraction of hydrogen in the local interstellar medium \citep{Nieva2012}.  In total, we have $\sim 665~M_\odot$ of gas within this region, based on the Herschel-derived column densities.

We determine a SFE for the Cosmic Cliffs region following the procedure of \citet{Evans2009}, i.e.,
\begin{equation}
    \mathrm{SFE} = \frac{M_*}{M_{gas}+M_*}. \label{eq:esf}
\end{equation}
 We then tabulate the SFE values in each of the aforementioned cases in Table~\ref{tab:sfrsfe}. Ranges for the values are determined by whether the number of YSOs was determined by nearest neighbor surface density (lower bound), or by counting the number of YSOs in each contour level based on their respective probabilities determined from the PRF method (upper bound).  

  {The inclusion of the sub-stellar cYSOs into the calculation of SFE has a relatively small impact. Though there are approximately three times as many sub-stellar YSOs as regular cYSOs identified in the Cosmic Cliffs, their contribution to the total $M_*$ is minimized given our assumption of their average mass being just 10\% that of the regular cYSOs.}

\begin{deluxetable}{cc}
    \tablecaption{Lower limits on the star formation efficiencies for two separate cases.
    \label{tab:sfrsfe}}
    \tablehead{\colhead{Case} & \colhead{SFE}}
       \startdata
        {No sub-stellar } & 0.08($\pm0.01$) - 0.09($\pm0.01$)\\
        {Mass correction}  & 0.08($\pm0.01$) - 0.10($\pm0.01$)\\
    \enddata
    \tablecomments{ In the first case, the sub-stellar population is removed from the sample. The second assumes that the sub-stellar component has a mean mass 10\% that of the regular YSOs.
    The range in each case is given by the surface density calculations (lower bound) and the number of cYSOs identified with the PRF (upper bound).}
\end{deluxetable}

 {Nevertheless, }the SFEs are  {significantly }higher than those found in previous cloud-level surveys \citep[$0.030-0.063$;][]{Evansc2d}.  {This increased SFE may largely be} due to eliminating the biases caused by the resolution and sensitivity limitations of previous instruments {, or it may also be due to the fact that we are calculating a SFE on a small site of triggered star formation, rather than over a whole cloud.}
Note that the values in both cases are still lower limits of the SFE, as the number of cYSOs we identify is limited by our training set. Hence, we expect that future JWST surveys will greatly change our understanding of the local efficiencies of star formation.
\section{Conclusions}\label{sec:conclude}

Star formation studies have thus far been limited by the resolutions and sensitivities of the instruments being used, being unable to sample the full IMF or obtain a clear census uncontaminated by blended sources. The James Webb Space Telescope far surpasses the limitations of earlier infrared space telescopes such as Spitzer, and provides rich fields of thousands of objects. The task we face now is determining how to identify in these newly rich fields the objects that sample the full range of the IMF that we wish to study further.
In this paper, we examined the utility of the Probabilistic Random Forest applied to JWST NIRCam data, which, based on the field of views of the respective detectors, will often cover much wider areas than MIRI data. We aimed to determine how well this machine learning method is able to identify new YSOs, based on a training set of known, Spitzer-identified YSOs within the field of a publicly available JWST dataset. 
Analysis of the new cYSOs and the metrics of our machine learning set yielded the following:

\begin{enumerate}
    \item \textbf{The PRF was able to recover Spitzer-identified YSOs to 98\% F1-Score.} Other metrics included an AUC score of 1.00, a Precision of 1.00, and a Recall of 0.96. These metrics indicate that the PRF was accurately able to identify most contaminants in its classifications.
    \item \textbf{In total, we retrieved \finaltotysos{} cYSOs from the entire set, including \finaltotnewysos{} cYSOs not previously identified in any work.} A number of embedded cYSOs identified in previous works are not recovered as cYSOs in our work due to the limitations of our training set, which does not well sample more embedded Class 0/I YSOs or more evolved Class III YSOs. 
    \item \textbf{JWST NIRCam data can be limited by extinction caused by dust, though to a far lesser extent than Spitzer data were (at wavelengths in common), as well as saturation of luminous and crowded YSOs.} The limitations of only using NIR data are that a number of x-ray sources and deeply embedded sources remained undetected due to intervening dust, while the saturation limits lead to several outflow sources being not measurable and thus unable to be classified. These limitations also matched that of our training set, leading to a bias in our cYSOs being preferentially less-embedded Class I and Class II sources.
    
    \item \textbf{The cYSOs closely follow the column density of gas in the region.} This spatial correlation is further supportive of previous works that found that Gum 31 is currently experiencing triggered star formation through the collect-and-collapse scenario.
    \item \textbf{The Cosmic Cliffs includes a substantial sub-stellar population.} This population is found in the HII region, and makes up $\sim$\fracsubs{} of our cYSOs.
    \item \textbf{The sub-stellar population calls into question previous estimates of the yields of star-forming molecular clouds.}  We determine SFEs that exclude and include this population and find them to be  {marginally} higher in the latter case  {due to our estimate of their mass being 10\% that of a regular cYSO}. 
    As a consequence, we expect future  {yields of cYSOs} in other regions using JWST data to be also higher than determined before, given this previously under-sampled very low-mass population. 
\end{enumerate}

We have found that the PRF method is able to quickly and accurately identify YSOs within JWST NIRCam data. Additionally, since the PRF method outputs the probabilities of an object being a YSO, it allows for Monte-Carlo sampling of the population of objects, and leaves open the option of changing the cYSO probability threshold as desired. Lastly, the PRF method is available in the popular coding language of Python, making it more accessible than other more complicated methods which are also able to classify objects with missing data, such as the Random Forest for Survival, Regression, and Classification in R \citep{RF_in_R2023}. 

Future work will focus on using data from multiple star-forming regions, particularly at a wider range of wavelengths. This will allow us to further fill out the training set and identify a wider range of YSO types, including more embedded Class 0/I YSOs. 

\begin{acknowledgements}
{The authors thank the anonymous referees for their constructive feedback, which have improved the quality of this paper.}
The authors would like to thank Peter Stetson, Chris Willott, and Nicholas Martis for their advice on JWST photometry.  In addition, we thank the staff of the NASA-funded Mikulski Archive for Space Telescopes (MAST) for providing the data products used for this paper.  JDF acknowledges the support of an NSERC Discovery Grant held at the University of Victoria.

\end{acknowledgements}

\appendix
\section{Outflow driving sources in the Cosmic Cliffs} \label{app:reit}
{In the following, we briefly describe the outflow driving source candidates determined in the Cosmic Cliffs field \citep{Reiter2022} matched to JWST objects we identify, give the PRF-based probabilities of these being YSOs, and suggest alternatives if we find these have less than 50\% probability of being a YSO. A label is assigned to each driving source candidate as being either FN = False Negative -- the object is very likely a YSO but the PRF algorithm classifies it as a contaminant, TP = True Positive -- the object is very likely a YSO and the PRF classifies it as such, MC = Misclassified by \citet{Reiter2022} -- we find a candidate with higher probability along the outflow axis, NP = New Positive -- a cYSO identified in our work that was not previously found, e.g., with Spitzer, or ND = No Detection -- the candidate driving source did not have sufficient good photometry to be included in the catalog.}

~

{MHO 1632: An arced outflow seen to either side of the suggested driving source, J103642.3-583804, which is located deepest within the cloud and whose SED does not broach 20 mags in the NIRCam bands. With MIRI data, this YSO may become more identifiable.  It has a probability of being a YSO of  {44\% , and is identified as a cYSO in 3 of 6 runs (where in each of these runs one of the filters is removed from the input data)}. No other likely candidates are found along the outflow axis, and as such this candidate is likely a false negative. (FN)}

{MHO 1633: An arced, sharp J-shaped outflow originating from J103648.0-583819. This object lies about midway within the dust cloud.  This object has probability $<1$\%, with no other possible driving sources along the outflow axis, and as such this candidate is likely a false negative. The object is most likely not identifiable as a YSO due to its faint SED. (FN)}

{MHO 1634: Several bow shocks are identified on either side of the driving source candidate, J103646.7-583805. This object has a probability of  {63}\% of being a YSO, with no other objects along the axis having greater than 10\% probability of being a YSO.  It was previously identified as a YSO in SPICY {, though not included in the training set due to its signal being overlapped with outflow signal}. (TP)}

{MHO 1635: An arced outflow on either side of the axis. This object is only visible in two bands (F444W and F470N) and as such we did not have enough data to classify it as a YSO or not. No other nearby objects along the outflow axis have greater than 10\% probability of being a YSO. \citet{Reiter2022} suggest that this object is only visible in MIRI data. Further studies will determine if we can retrieve it using the PRF method. (ND)}

{MHO 1636: a series of three separate outflows, which can be traced back towards an object which lies midway within the dust cloud, candidate driving source J103651.5-583754. The PRF yields a probability of  {54\%, though it is only classified as such in 3 of 6 runs.} 
(FN)}

{MHO 1637: Bipolar outflows originating from J103650.5-583752, an object with $\sim 100\%$ probability of being a YSO, which was previously identified in Spitzer data and was part of the training set. (TP)}

{MHO 1638: A series of several outflows, tracing a gentle c-shape, with the driving source candidate in the middle of the c appearing to originate from J103651.4-583748. This object has only a 1\% probability of being a YSO, and all others along the outflow axis have a probability less than that. The nearest YSO candidate along this axis is J103652.3-583809, which is also associated with HH 1219, has a probability of $\sim 100$\%, and was not previously identified in Spitzer data.  By visual inspection, the HH outflow also appears to be along the same axis. (MC, NP)}

{MHO 1639, HH 1221, HH 1003  Aa: Driving source J103653.8-583748 has a probability of being a YSO of  {78}\%, but was not previously identified as a YSO in Spitzer.  The outflows trace a bright object located on the cusp of the cloud edge. This object is hidden in the shorter wavelengths, and appears to light up the edge of a nearby cavity, see \citet{Reiter2022} for details. (NP)}

{MHO 1640: Asymmetric outflow appearing to originate from J103651.5-583710, which has a probability of being a YSO of  {51\% with the PRF, though is only classified as such in 3 of 6 additional runs}. Slightly further along the outflow axis, which would lead to a more symmetric outflow, is a fainter driving source candidate with a probability of being a YSO of  {87\%}, J103651.3-583709. (MC, NP)}

{MHO 1643, HH 1218: The clear outflow driving source here J103654.2-583626 has twin bow shocks but was unable to be classified by us as a YSO due to saturation at long wavelengths as well as being too faint in short wavelengths. Indicative of the shortcomings of retrieving photometry from JWST images. (ND)}

{MHO 1645, MHO 1646: Bipolar outflow and H$_2$ knots trace the possible driving source J103654.4-583618, which we find has 66\% probability of being a YSO.  (NP)}

{MHO 1647, HH 1002a: Two bow shocks with proper motion trace driving source J103654.0-583720; this object has a $\sim 99\%$ probability of being a YSO and was found with Spitzer. (TP)}

{MHO 1649: A series of four nearby shocks trace J103653.6-583520 which has probability of being a YSO of 6\%.  No other possible driving sources are seen along the outflow axis. (FN)}

{MHO 1650: Single outflow with suggested driving source J103653.1-583737 (PRF probability $<$1\% of being a YSO), which is coincident with another bright spot of H$_2$ emission. Another possible driving source,  {J103652.9-583737}, however, lies along the outflow axis with a probability of being a YSO of  {70\%}. (MC, NP)}

{MHO 1651, HH 1003 B, Ca: Two knots originate from driving source J103653.3-583754, which has a probability of being a YSO of $\sim 100\%$ and was previously identified in Spitzer data and hence was part of the training data. (TP)}

MHO 1652:  Two knots trace driving source J103652.7-583805, which has a probability of being a YSO of  {67\%. (NP)}

All HH objects had proper motions and could be traced back to their source.

HH c-3: Driving source J103701.5-583751 has probability  {$40\%$ (PRF and RF)}. { No other sources lie along the outflow axis. (FN)}

{HH c-4: Driving source J103702.1-583658 has probability $\sim 100\%$ and was previously identified in Spitzer data and hence was part of the training set. (TP)}

{HH c-5: Possible driving source J103653.9-583632 does not have sufficient photometric data available ($>2$ bands). Nearby source J103653.7-583632 has probability 27\% (PRF) or 41\% (RF), respectively. (ND)}

HH 1219: Possible driving source J103652.3-583809 has probability $\sim 100\%$ and was previously identified in Spitzer data and hence was part of the training set. (TP)

{HH 1223a: Clear driving source is too saturated to obtain classification, but was previously identified with Spitzer. (ND)}

{We note that false negatives generally seem to be attributed to faint sources, while non-detections are attributed to over-saturated sources. With MIRI data included in the ensemble, we may be able to recover many as cYSOs. In total, we find 5 False Negatives, 6 True Positives, 3 Misclassifications, 6 New Positives, and 4 Non-Detections.}
\bibliography{main.bib}
\end{document}